\documentclass[a4paper,11pt]{article} \pdfoutput=1 

\usepackage{jinstpub} 

\title{\boldmath 
The FAMU experiment  at RAL
to study the  muon transfer rate
from hydrogen to other gases}

\author[a]{A.~Adamczak,}
\author[b,r]{G.~Baccolo,}
\author[b,q]{S.~Banfi,}
\author[c]{D.~Bakalov,}
\author[d]{G.~Baldazzi,}
\author[b,r]{R.~Benocci,}
\author[b]{R.~Bertoni,}
\author[b,q,1]{M.~Bonesini, \note{corresponding author}}
\author[e]{V.~Bonvicini,}
\author[b]{F.~Chignoli,}
\author[b,q]{M.~Clemenza,}
\author[h,i]{L.~Colace,}
\author[e,u]{M.~Danailov,}
\author[c]{P.~Danev,}
\author[f,g]{A.~de Bari,}
\author[g]{C.~De~Vecchi,}
\author[h,j]{M.~De~Vincenzi,}
\author[e,p]{E.~Furlanetto,}
\author[d,x]{F.~Fuschino,}
\author[e,k,t]{K.~S.~Gadejisso-Tossou,}
\author[e,w]{D.~Guffanti,}
\author[h]{A.~Iaciofano,}
\author[l]{K.~Ishida,}
\author[d,x]{C.~Labanti,}
\author[b,r]{V.~Maggi,}
\author[d]{A.~Margotti,}
\author[b]{R.~Mazza,}
\author[f,g]{A.~Menegolli,}
\author[e]{E.~Mocchiutti,}
\author[b,r]{M.~Moretti,}
\author[d,x]{G.~Morgante,}
\author[b,q]{M.~Nastasi,}
\author[k]{J.~Niemela,}
\author[e]{C.~Pizzolotto,}
\author[b,q]{E.~Previtali,}
\author[m,n]{A.~Pullia,}
\author[m,s]{R.~Ramponi,}
\author[e]{A.~Rachevski,}
\author[d]{L.~P.~Rignanese,}
\author[g]{M.~Rossella,}
\author[p]{N.~Rossi,}
\author[e,y]{R.~Sarkar,}
\author[c]{M.~Stoilov,}
\author[e,k]{L.~Stoychev,}
\author[g,o]{A.~Tomaselli,}
\author[h]{L.~Tortora,}
\author[e]{E.~Vallazza,}
\author[e]{G.~Zampa }
\author[e,l,p]{ and A.~Vacchi,}
\newpage
\affiliation[a]{Institute of Nuclear Physics, Polish Academy of Sciences, Radzikowskiego 152, PL31342 Krak\'{o}w, Poland}
\affiliation[b]{Sezione INFN di Milano Bicocca, Piazza della Scienza 3,  
Milano, Italy}
\affiliation[c]{Institute for Nuclear Research and Nuclear Energy, Bulgarian Academy of Sciences, blvd. Tsarigradsko ch. 72, Sofia 1142, Bulgaria}
\affiliation[d]{Sezione INFN di Bologna, viale Berti Pichat 6/2,  Bologna,
Italy}
\affiliation[e]{Sezione INFN di Trieste, via A. Valerio 2,  Trieste, Italy}
\affiliation[f]{Dipartimento di Fisica, Universit\`a di Pavia, via Bassi 6, 
 Pavia, Italy}
\affiliation[g]{Sezione INFN di Pavia, Via A.~Bassi 6,  Pavia, Italy}
\affiliation[h]{Sezione INFN di Roma Tre, Via della Vasca Navale 84, 
Roma, Italy}
\affiliation[i]{Dipartimento di Ingegneria, Universit\`a degli Studi Roma Tre,
 Via V. Volterra 62,  Roma, Italy}
\affiliation[j]{Dipartimento di Matematica e Fisica, Universit\`a di Roma Tre, 
Via della Vasca Navale 84, Roma, Italy}
\affiliation[k]{The Abdus Salam International Centre for Theoretical
Physics, Strada Costiera 11, Trieste, Italy}
\affiliation[l]{Riken Nishina Center, RIKEN, 2-1 Hirosawa, Wako, Saitama 351-0198, Japan}
\affiliation[m]{Sezione INFN di Milano, via Celoria 16, Milano, Italy}
\affiliation[n]{Dipartimento di Fisica, Universit\`a degli Studi di Milano,
via Celoria 16, Milano, Italy}
\affiliation[o]{Dipartimento di Ingegneria, Universit\`a di Pavia, 
Via Ferrata 5, Pavia, Italy}
\affiliation[p]{Dipartimento di Scienze Matematiche, Informatiche e Fisiche, 
Universit\`a di Udine,
  via delle Scienze 206, Udine, Italy}
\affiliation[q] {Dipartimento di Fisica G. Occhialini, Universit\`a di Milano
Bicocca, Piazza Scienza 3, Milano , Italy}
\affiliation[r]{Dipartimento di Scienze dell'Ambiente e della Terra, 
Universit\`a di Milano Bicocca, Piazza Scienza 1, 20126 Milano, Italy}
\affiliation[s]{IFN-CNR, Dipartimento di Fisica, Politecnico di Milano,
piazza Leonardo da Vinci 32, Milano, Italy}
\affiliation[t]{Laboratoire de Physique des Composants \`a Semi-conducteurs
(LPCS), Department de physique, Universit\'e de Lom\'e, Lom\'e,
Togo}
\affiliation[u]{Sincrotrone Elettra Trieste, SS14, km 163.5, Basovizza,
Italy}
\affiliation[w]{Gran Sasso Science Institute, via F. Crispi 7, L' Aquila, Italy}
\affiliation[x]{INAF-IASF Bologna, Area della Ricerca, via P.~Gobetti 101, 
             Bologna, Italy.}
\affiliation[y]{Indian Centre for Space Physics, 43
    Chalantika, Garia Station Road, Garia, Kolkata, 700084 West
    Bengal, India}
\emailAdd{maurizio.bonesini@mib.infn.it}

\abstract{
The  aim  of the FAMU  
(\underline{F}isica degli \underline{A}tomi \underline{Mu}onici) 
experiment is to realize the first measurement of the 
hyperfine splitting (hfs) in the 1S state of muonic
 hydrogen $\Delta E^{hfs}_{1S}$, 
 by using the RIKEN-RAL intense pulsed muon beam 
and a high-energy mid-infrared tunable laser. 
This requires a detailed study of the 
muon transfer mechanism at different 
temperatures and hence at different epithermal states of the muonic system. 
The experimental setup  involves a cryogenic pressurized gas target and a 
detection system based on silicon photomultipliers-fiber beam hodoscopes
and  high purity Germanium detectors and Cerium doped Lanthanium Bromide 
crystals, for X-rays detection at energies around 100 keV. 
 
Simulation, construction and  detector performances of the FAMU 
apparatus at RAL are reported in this paper.

}


\collaboration{
\includegraphics[height=17mm]{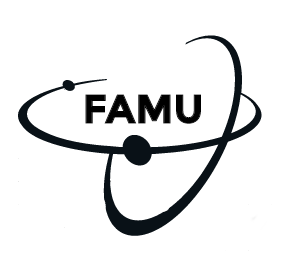}
 FAMU collaboration}

\newpage

\usepackage{lineno}
\begin{document}
\maketitle
\section{Introduction
}

The FAMU proposal \cite{vacchi2012,adamczak2012}  
aims to a new high precision measurement of muonic Hydrogen spectroscopy.
In this experiment, incident low momentum muons from pulsed muon source are stopped in a 
Hydrogen gas mixture at appropriate temperature and pressure  to form
 muonic Hydrogen atoms $\mu p$. 
Once the thermal $\mu p$  have reached the ground para (F=0) state, 
they are exposed to photons of energy equal to the hyperfine-splitting 
resonance-energy $\Delta E^{hfs} \simeq$~0.182 eV, and excited
to the ortho (F=1) spin state. They then very quickly de-excite to the
(F=0) state in subsequent collisions with the surrounding 
$H_2$  molecules~\cite{adamczak96}. 
 At the exit of the collision the muonic atom is accelerated, because of energy and momentum conservation, by ~2/3 of 
the excitation energy $\Delta E^{hfs}$. 

In the following thermal collisions 
of $\mu p$, part of the muons are transferred to a low concentration 
admixed gas Z, forming excited states of $\mu Z$. The rate of this reaction depends 
on the kinetic energy of the muonic atoms.

\begin{figure}
\centering
\includegraphics[width=.99\textwidth]{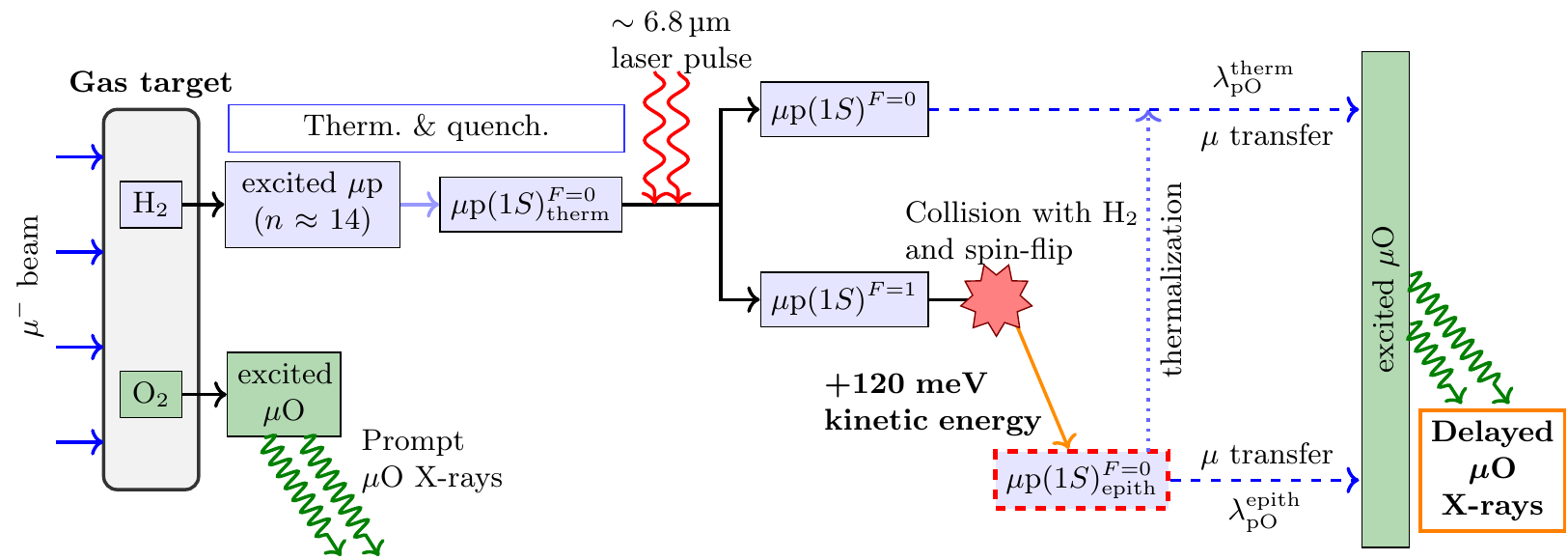}
\caption{Schematic representation of the employed experimental method.
Here Oxygen is the admixed gas Z. }
\label{fig:1}
\end{figure}

By tuning the emission wavelength 
of the laser around $\Delta E^{hfs} \simeq$~0.182 eV, corresponding to
6.78~$\mu$m, it is possible to experimentally determine the number of 
muonic atoms that have undergone the above sequence of processes, and 
identify the resonance wavelength as the value for which the number of 
spin-excited atoms is maximal. As outlined in figure~\ref{fig:1}, the muonic 
Hydrogen atoms are created and then
propagate in a gas  target with an appropriate 
mixture of Hydrogen and a higher-Z contaminant. 
The observable is the time distribution of the characteristic X-ray emitted
from the muonic atoms produced by muon transfer from Hydrogen to the atom of 
the admixture gas $ (\mu p) + Z \rightarrow  \mu Z^{*} + p$  and its response 
to variations of the laser excitation wavelength \cite{bakalov92,bakalov15} . 
These X-rays are distinguished by the prompt X-ray from muonic 
atoms, formed in direct prompt capture, by the time delay with respect 
to the beam arrival. 
The $(\mu^{-} p)_{1S} $  hfs resonance is recognized by the maximal response 
to the tuned laser wavelength of the time distribution of X-ray K-lines 
from $(\mu Z)^{*}$  cascade, i.e. by the maximal difference between 
the time distributions in presence and without laser radiation. 

\section{The FAMU  experimental setup }
The pulsed muon source of the RIKEN 
 muon facility at the Rutherford 
Appleton Laboratory (RAL) \cite{matsuzaki01}, 
is well suited to the purpose of the project. It can deliver about 
3 to 8~$ \times 10^4$  negative muons per second with a pulse repetition 
rate of 50~Hz and 
momentum in the range 30--80 MeV/c, with $\sigma_{p}/p= 4 \%$ and a 
beam transverse r.m.s  $\sigma_x,\sigma_y = \ 1.5$~cm.
The beam has a double peak structure with 70 ns pulse width (FWHM) 
and peak to peak distance of 320 ns. The beam can be delivered 
alternatively to four experimental ports. 

The realization
 of the FAMU project has been planned into several steps. The first step
took data in June 2014~\cite{adamczack14},
with a preliminary setup 
and allowed to publish  first  results on muon transfer
rate to the heavier admixed gas \cite{vacchi16,mocchiutti17}.
It has been followed by the experiment described in 
detail in this paper, that took data from 2016 on. The setup with the laser will be implemented in a future phase.
\begin{figure}
\centering
\includegraphics[width=.99\textwidth]{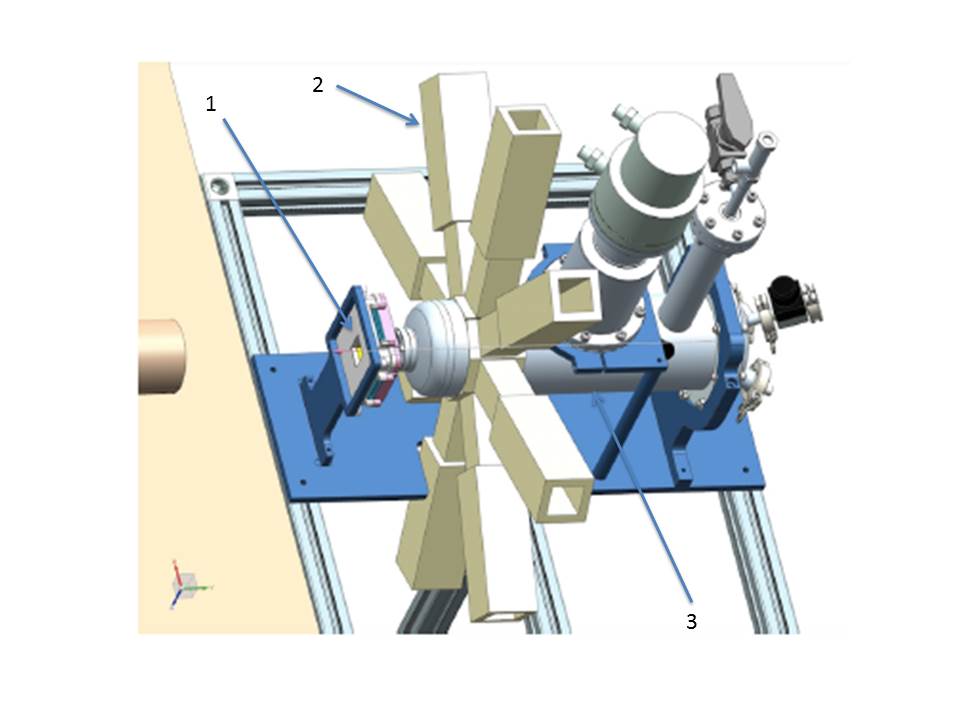}
\caption{Lateral artist view of the FAMU experiment
 at RIKEN-RAL, as in 2016 data taking. The beam comes from 
the left. Visible are the 1 mm pitch beam hodoscope (1), 
the Ce:LaBr$_3$  crystals with photomultiplier (PMT) readout (2), arranged in a crown of 
8 elements and the cryogenic target (3). 
The HPGe detectors are not drawn in this view.}
\label{fig:det}
\end{figure}
The experimental apparatus of the FAMU experiment is based on: 
\begin{itemize}
\item{} a 1 mm pitch 
X/Y beam hodoscope to tune  the beam steering onto the target and 
evaluate the muon incoming rate, thus optimizing 
the efficiency of the data taking;
\item{} a cryogenic target, where Hydrogen gas admixed with an 
heavier contaminant is contained;
\item{} a system for characteristic X-rays detection, based mainly 
on 1" thick cylindrical Cerium doped Lanthanium Bromide (Ce:LaBr$_3$) crystals and 
High Purity Germanium (HPGe) detectors.
\end{itemize}
 A schematic layout is shown in figure \ref{fig:det}.
The apparatus described here was used for four data taking periods, 
mainly in February 2016. 
In detail, the  main goals of the 2016 data taking were:
\begin{itemize}
\item{} measure the muon transfer rate from muonic Hydrogen to Oxygen and other
gases at different temperatures (between 300 and 100~K), pressures
and concentration;
\item{} validate the theoretical calculation about the best Oxygen 
concentration;
\item{}  study different contaminants which could give a temperature (energy) 
         muon transfer rate dependence as Oxygen;
\item{}  perform background tests with empty target or filled
with  pure Hydrogen or pure Nitrogen.
\end{itemize}
The collected runs, taken in February 2016,  are resumed in table \ref{tab1}.
\begin{table}[htb]
\caption{Data sets taken in the 2016 FAMU experiment at RIKEN-RAL. All
runs were taken at 57 MeV/c beam momentum.}
\label{tab1}
\smallskip
\centering
\scalebox{0.8}{%
\begin{tabular}{|l|c|c|c|c|}
\hline
gas mixture   &  temperature       & data taking   & goals &filling  at \\
              &  (K)               &               &  & \\
\hline
\hline
$N_2$ (100\%)    & 298                      & 6 h           & det.
calib/test  &   2 bar 300 K \\
\hline
$H_2 + O_2$ (0.3\%)  & 300/273/240/ & 3 h/step  & thermal cycle   
               &  41 bar 300 K \\
            &   200/150/100              &                   & & \\
\hline
$H_2 + O_2$ (0.3\%)  & 100                      & 20 m        &  high p/low T test         &   30 bar 100 K \\
\hline
$H_2 + O_2$ (0.05\%) & 300/273/240              & 3 h/step & thermal cycle   
               &   41 bar 300 K \\
\hline
$H_2 + Ar$ (0.3\%)   & 300/273/240/ & 3 h/step & thermal cycle   
               &   41 bar 300 K \\
             &             200/150/100      &          &        & \\
\hline
Vacuum            & -                         & 3 h           & backg studies          & $\approx 10^{-6}$ mbar \\
\hline
$H_2 + Ar$ (1\%)     & 300/273/240              & 3 h/step & thermal cycle   
               &   41 bar 300 K \\
\hline
$H_2 + O_2$ (1\%)    & 300/273/240              & 3 h/step & thermal cycle   
               &   41 bar 300 K \\
\hline
$H_2$ (100\%)    & 300                      & 30 m        & backg studies          &   41 bar 300 K \\
\hline
$H_2 + O_2$ (0.3\%) & 300                      & 30 m        & 
check 2014 results &   41 bar 300 K \\
\hline
$H_2 + CH_4$ (0.3\%) & 300/273/240/        & 3 h/step & thermal cycle   
               &   41 bar 300 K \\
                  & 200                & &  & \\
\hline

\end{tabular}}
\end{table}
\subsection{The 1 mm pitch beam hodoscope}
A relevant  issue for the FAMU experiment is the optimal steering of the
high intensity pulsed muon beam impinging onto the Hydrogen target, to
maximize the muonic Hydrogen production rate.
A system of three beam hodoscopes based on square 
scintillating fibers read by silicon photomultipliers (SiPM) 
has been developed for this scope: two have a 3~mm pitch and are removable
and are used only for special runs~\cite{carbone14}, 
while the last has a  1~mm pitch
and is permanently installed in front of the target entrance~\cite{bonesini17}.

\begin{figure}[htbp]
\centering 
\includegraphics[width=.9\textwidth]{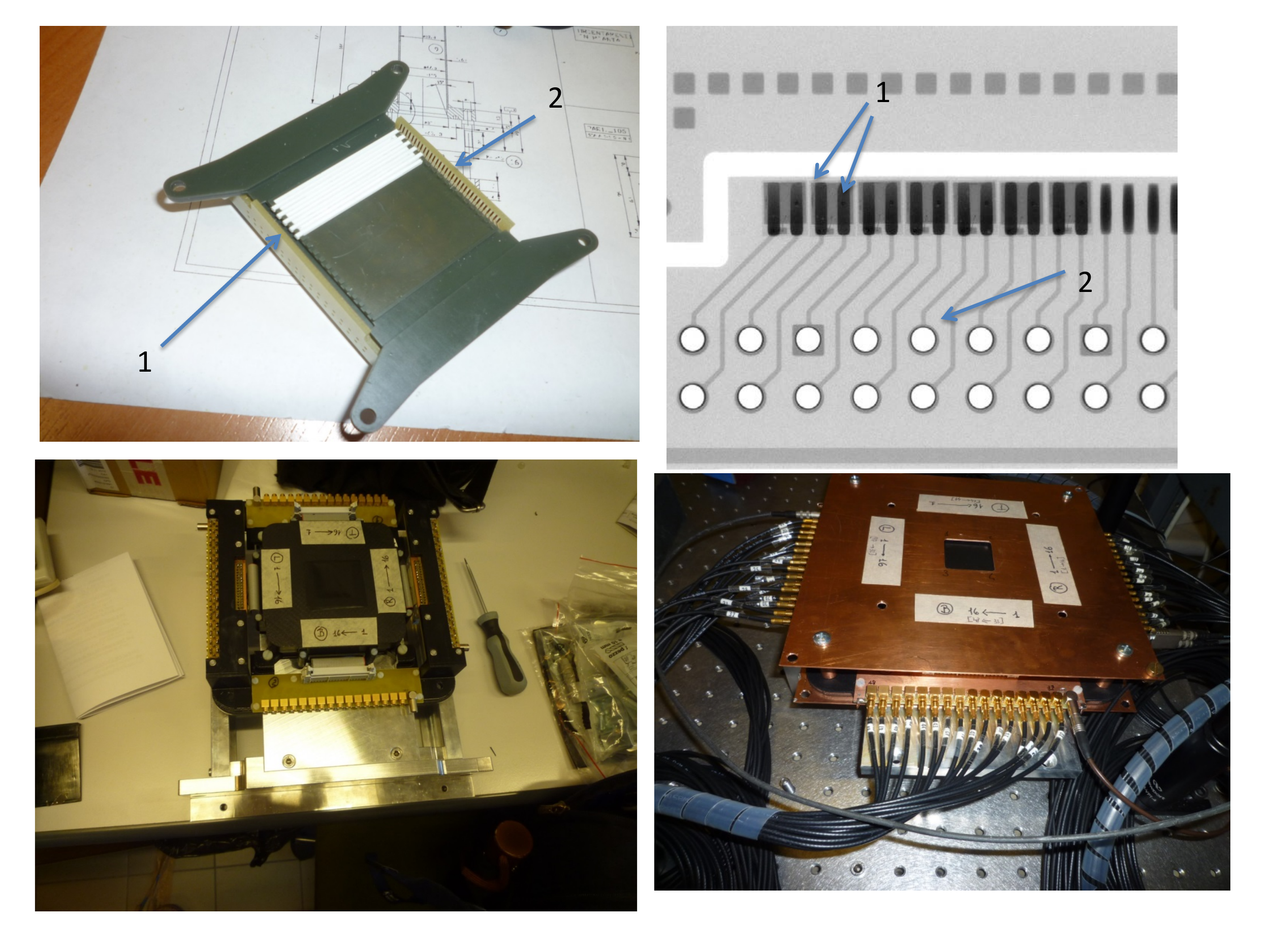}
\caption{Left-top panel: holder of the 1 mm square fibers. Visible are some
         white coated scintillating fibers (1) and the printed circuit
board (PCB) (2), where SiPM are soldered.
Right top panel: X-rays image to cross-check the SiPMs mounting on the PCB.
(1) are the conductive tracks, where cathode and anode of a SiPM are soldered; (2) is 
the footprint where the 40-way connector for flat cable is mounted. 
Left-bottom panel: mounted detector, with an 
interface board visible. Signals cables are to be attached to the row of MCX 
connectors. Right-bottom panel: complete detector, with two 1.5 mm copper plates
for electrical shielding.}
\label{fig:ass2}
\end{figure}

To reduce the amount of material in front of the target entrance window,  
1 mm$^2$ square scintillating Bicron BCF12 
fibers coated with white EMA, to avoid light cross-talk, were 
used. The fibers (32+32)  were arranged in parallel on two orthogonal 
planes along X/Y coordinates,
giving a detector active area of $32 \times 32$~mm$^2$. 
RGB SiPMs from Advansid, with a total $1 \times 1$~mm$^2$ active area 
and 40~$\mu$m cells,
were used to detect scintillation light. 
Because the SiPM's footprint is slightly bigger than
the fiber cross-section, fibers had to be read alternating left/right and 
up/down sides. 
All available SiPMs  were tested individually to determine
their breakdown voltages by measuring their current-voltage characteristic 
curve. 
It was 
 possible, with a suitable
selection of the SiPMs to be used, to employ a common voltage for the 
biasing of the SiPMs of each detector side. 
The SiPM's  signals are then fed into a 
5~Gb/s CAEN V1742  FADC and  processed via the  FAMU DAQ system based on a CAEN V2718 
VME-PCI interface. 
Onboard 100 k$\Omega$ SMD thermistors have been mounted to ensure
long-term stability via offline correction of the thermal drift of SiPMs' gain.
Some details of the detector mounting are shown in figure \ref{fig:ass2}.

\subsection{The cryogenic  target system}
\label{sec:targetsystem}
The efficiency of the method proposed by the FAMU collaboration
depends on the collisional energy dependence 
of the muon transfer rate. While for many gases the muon transfer rate from 
the $ \mu p$ system at low energies is nearly constant, there is 
experimental evidence  that for Oxygen there is instead a sharp energy 
dependence \cite{werthmuller98}. Monte Carlo simulations, based on these data, 
have shown that the method proposed by FAMU may provide the expected results 
\cite{bakalov15}. 
Its experimental verification  requires a detailed study of the muon transfer mechanism at different 
temperatures and hence at different epithermal state of the muonic system. 

The cryogenic gas target system was developed for this task with 
the following characteristics: 
\begin{itemize}
\item {}low-mass beam entrance window for minimal losses and lateral spread of 
the impinging muon beam;
\item{} high transparency of the target lateral walls to the X-rays of the 
muonic lines of interest;
\item{} capability to held  pressures up to 40 atm  
of ultrapure Hydrogen gas; 
\item{} ability to work at stable temperatures from  50 to
    300 K.
\end{itemize}
\begin{figure}
\centering
\includegraphics[width=.9\textwidth]{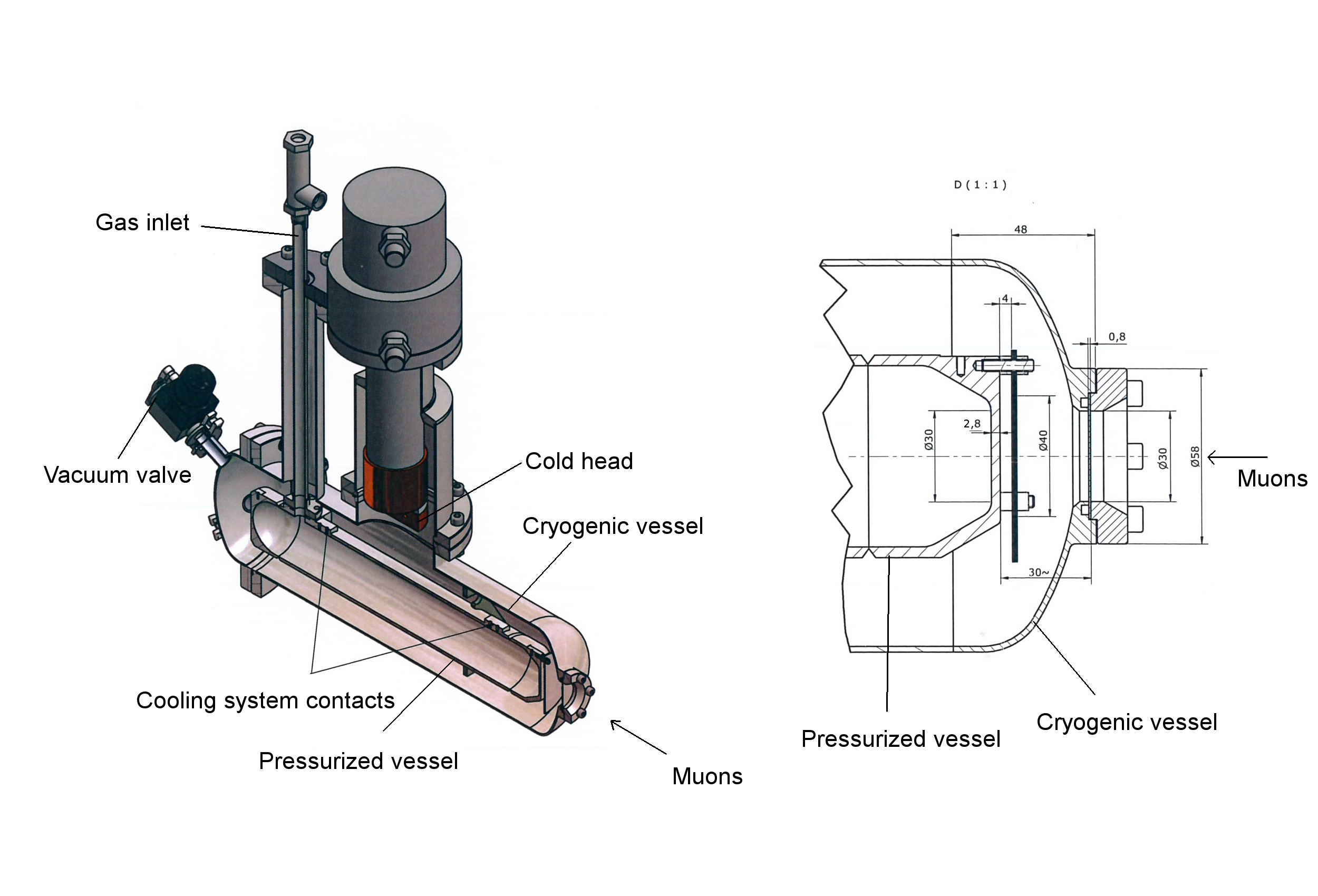}
\caption{The target design.
Left panel: the double volume shell needed for thermal insolation is clearly 
visible: the whole is realized in Aluminium alloy.  Right panel: the details of 
the muon beam entrance window. Beam goes from right to left.}
\label{fig:2}
\end{figure}
The resulting target 
design\footnote{Realized by Criotec Impianti srl \cite{criotec}.}  is shown in the left 
panel of figure \ref{fig:2}. 
An  Aluminium alloy cryostat insulates the internal target, containing the
gas, from the environment by means of a vacuum jacket, MLI superinsulation 
shrouds and fiberglass stunts.  In the left panel of figure~\ref{fig:2}, 
from right to left  muons will cross the first thin 
0.8~mm Aluminium entrance window entering  the evacuated volume, where a stack of
three Aluminium disks (0.1~mm each) with fiberglass ring spacers, are used 
as radiative shields,
before entering the pressurized volume through a second 2.8 mm Aluminium
 entrance 
window. The details of the muon 
beam entrance window are visible in the right panel of figure~\ref{fig:2}.  
A careful evaluation of the X-rays background contribution coming from 
the target has shown the importance of having a thin coating 
of the internal shell of the target 
made with high Z material (100~$\mu$m of Ni and 20~$\mu$m of Au). This allows fast nuclear capture of the muons reaching 
the walls, thus reducing the background coming from electrons from  muon decay.
The cryogenic system has been built and certified 
to comply with the standard EU safety rules. 

The cooling system is built around a Sumitomo CH-104 cold head, coupled
to a HC-4E1 helium compressor. It is a single-stage cryogenic 
refrigerator that operates on
the Gifford-McMahon refrigeration cycle, able to provide more than 20~W at a
temperature of 40~K. 
Water needed for heat extraction at the compressor is supplied 
by a closed system
water chiller with monitored temperature. During the whole data taking the water
temperature remained in the range from 8 to 20 degrees Celsius, well
within helium compressor specifications.

Four DT-670 Silicon diodes thermometers are integrated in the system: 
two on the cold head
and two at the ends of the inner cylinder. A Lakeshore 336 temperature 
controller is used to  control the target
temperature. 

Vacuum performances are very good: the system reaches a vacuum level
in the $10^{-5}--10^{-6}$~mbar range. Even once the thermo-vacuum
pump is disconnected (at low temperature) the vacuum level is
maintained for several days.

\subsubsection{Cooling system}

The measure of the muon transfer rate  requires
an efficient operation of the target at various temperatures, with a 
stability better than $\approx0.5 ^{\circ}$~C and with 
different gas
fillings. 
%
The  cold head reaches 28~K in about 3 hours and stabilizes to 
its steady-state equilibrium of 27.7~K in 4--4.5 hours. 
The target follows this trend with a time delay of nearly 40 minutes
due to its thermal capacitance.
In this ``steady-state'', the target reaches 30.7~K and 
 the stability is very good: around 0.01~K/h.  

The best approach to save time and have a good temperature stability
during system cooling is to perform temperature steps
not exceeding $\approx$50~K.
With this procedure, a conservative 1\% temperature variation at the
required set-point can be used as
the threshold for starting the measurement. At the 1\% level we can assume 
that the transfer rate from muonic Hydrogen to
other gases is almost constant. However, being the temperature
recorded in the raw data structure, any anomalous effect can be tracked down
 during the following data analysis. 
Operatively a set-point is reached when the average temperature of the
target is within 1\% of its value and the stability is better than 1\%
over 1000~s. The study of transfer rate temperature dependence requires to fix several target
temperatures. However it is not needed to set very precisely a  given
temperature. Thus thermal control can be achieved by picking a set-point
value and letting the target stabilize at its natural equilibrium
temperature. 

 Lowering the target temperature relies on the refrigerator
cooldown capability while heating is based on the power supplied by
the Lakeshore controller (0 -- 100~W with the presently installed resistor of
25~$\Omega$).
The cooldown rate, as expected, is not constant and it is inversely
proportional to the temperature.

\begin{figure}[!htb]
\centering
\includegraphics[width=0.9\textwidth]{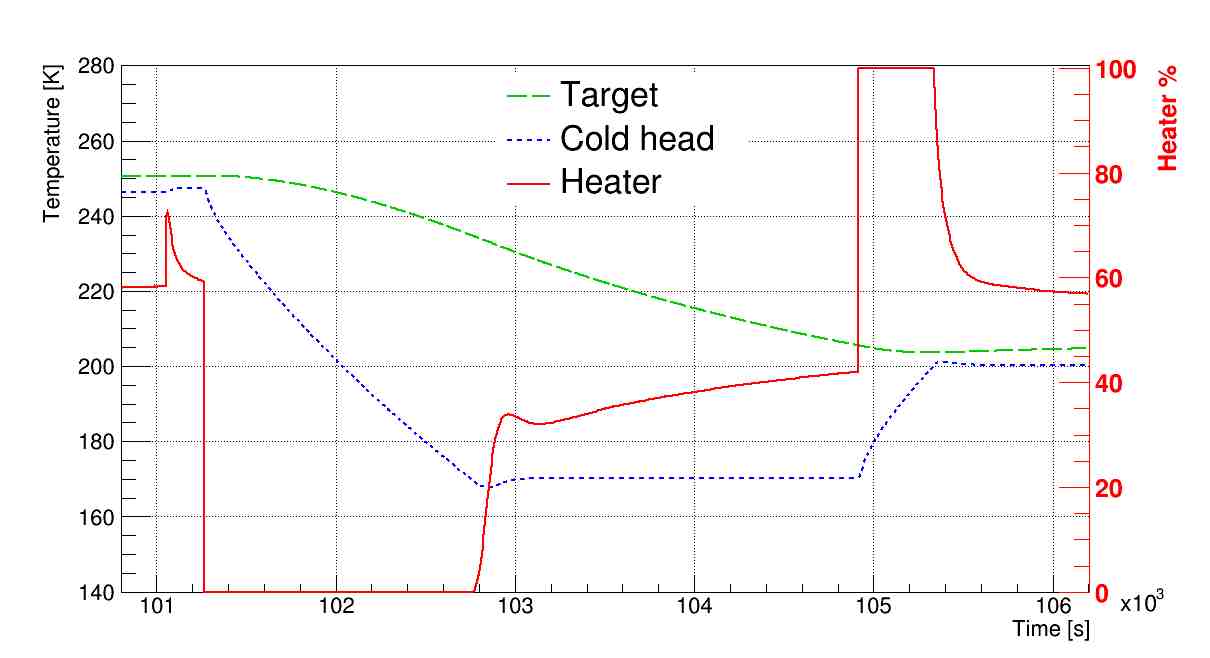}
\caption{Cooldown step from 250 K to 205 K. Temperatures of target and
cold head are shown (left scale) together with the heater power
(right scale) as a function of the elapsed time.}
\label{figT}       
\end{figure}

Figure~\ref{figT} shows the target and cold head temperature as
function of time during a laboratory test of the cryogenic system. The
target is cooled from 250~K to 205~K.
Even if, in this case, the control operations have not been optimized 
the time needed to cool down the target by $\approx$50~K is of about
one hour.

Thermal operations on the target must also take into account that
different gas mixtures and pressure are to be used. After each cycle
the target needs to be correctly evacuated and cleaned. 
The gas mixture re-loading process has to be done at
room temperature, as low temperature condensation and cryo-sorption issues may prevent
to reach the required level of cleanliness. Thus, each run with a new gas 
mixture follows a cooldown profile in steps of 50 K starting from room
temperature (from 300~K to 100~K), as described later in Section~\ref{target_op}.

\subsection{The Ce:LaBr$_3$ X-ray detectors with PMT readout}
The study of muon transfer rate time dependence requires a high speed and
high resolution X-ray detection system.
A fast detector system based on  Ce:LaBr$_3$
 1" scintillation crystals was thus developed.
Their high light output (63,000 photons/MeV at 380~nm) and 
fast  pulse time decay (16~ns) allow for high counting rate, excellent
energy and time resolution \cite{LABR1}.
The high light output and the high absorption efficiency
for X-rays (85\% at 300~keV for 1" diameter crystal \cite{sgobain_calc}) allow
very good spectroscopic performances (less than 3\% of FWHM  at
662~keV \cite{LABR1, LABR2}) when coupled with high quantum efficiency last generation photomultipliers (PMTs).

\begin{figure}
\centering
\includegraphics[width=.42\textwidth]{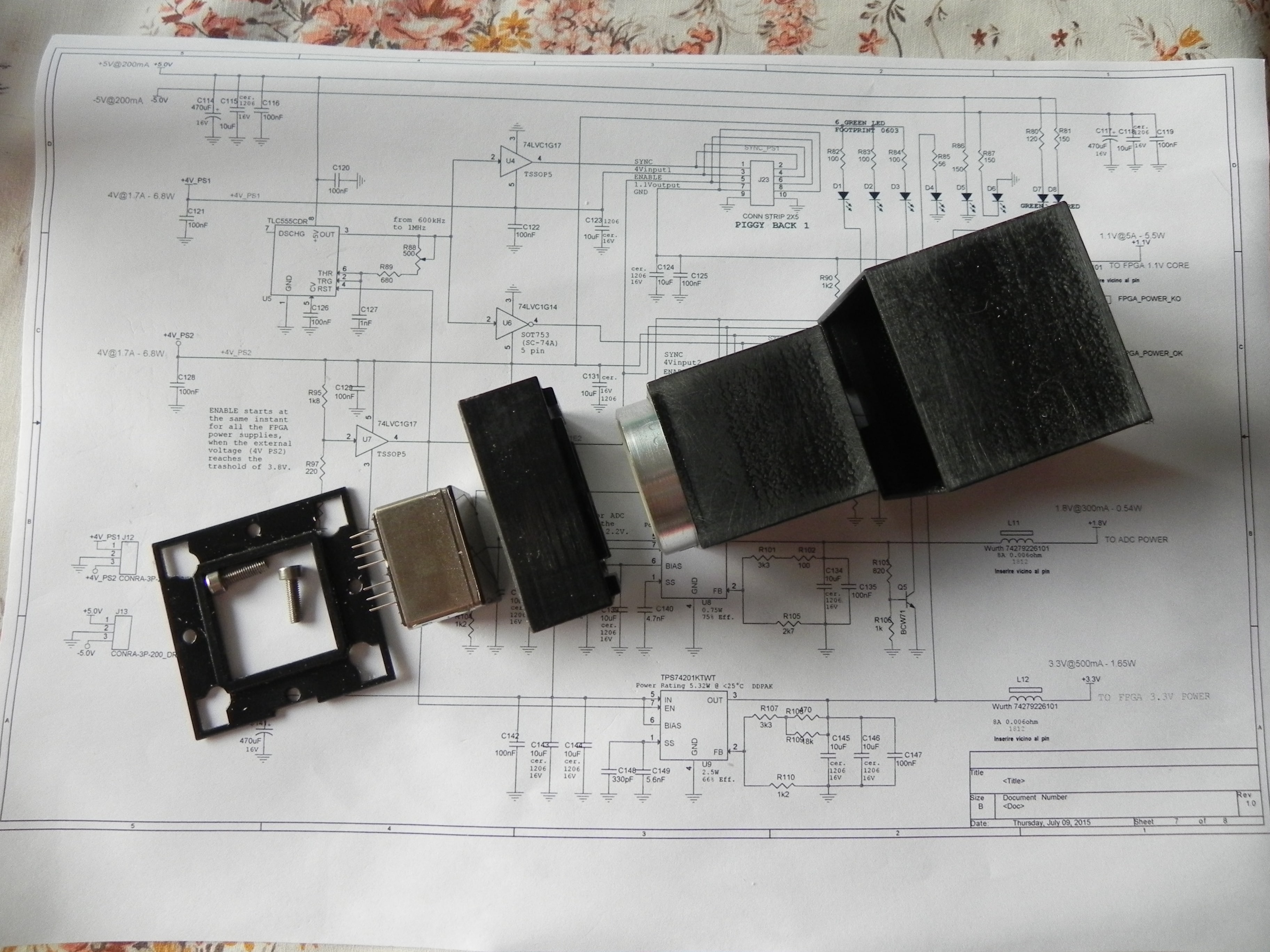}
\includegraphics[width=.42\textwidth]{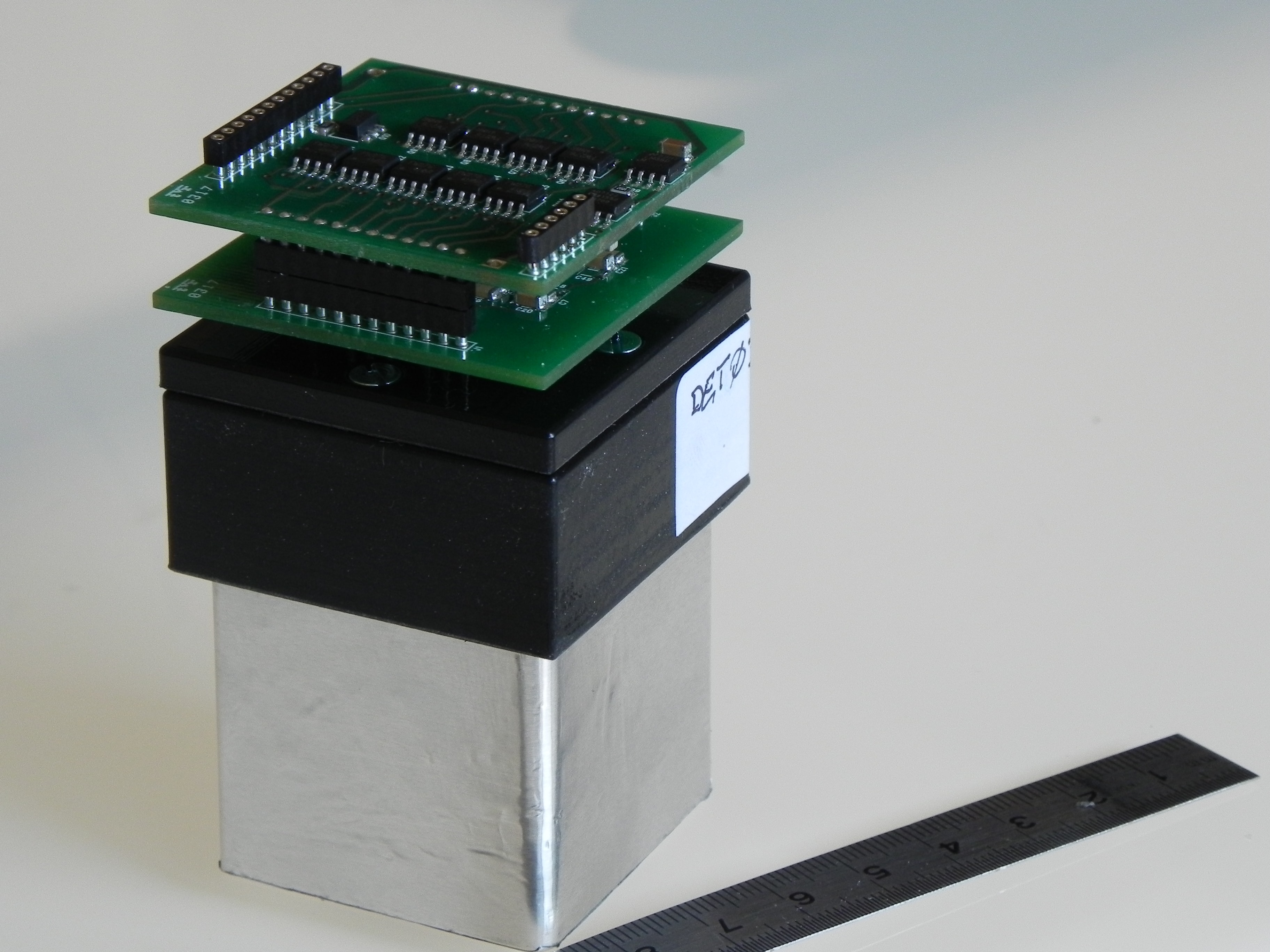}
\caption{Left panel: disassembled detector: the circular Ce:LaBr$_3$ crystal
and the squared PMT are visible. Right panel:  assembled detector with
 part of the
electronic visible, in particular the active voltage divider MOSFETs.}
\label{fig:PMT}
\end{figure}

To maximize the collection of X-rays produced by the decaying muonic atom cascade, 
a set of eight detectors was arranged as close as possible to the external 
target shell.
Taking into consideration
 the volume density of muons stopped in gas and the isotropic X-ray emission, 
the detectors  are arranged in star shaped
mechanical support (figure~\ref{fig:det}).
All detectors are placed 3~cm away from the internal target 
shell containing the gas mixture. Detectors cover a total area of 
40.5~cm$^2$ which correspond to 4.5~sr of solid angle. 
Each detector unit is composed by a 1"x1" 
(diameter x thickness) Ce:LaBr$_3$ crystal coupled with a compact 
PMT. Crystal, PMT and the electronics are encapsulated 
in a custom, black ABS 3D printed holder which also provides light tightness 
(see figure~\ref{fig:PMT}). 
The integrated detector is inserted in 
a 3~x~3~cm$^2$ 2~mm thick Aluminium  profile. 
Crystals with similar performances were selected from three manufacturers: Scitilion
(China), REXON (USA) and Saint Gobain (France). The 
PMTs are Ultra
BiAlcali (UBA) cathode photomultipliers  R11265U-200 
by Hamamatsu Photonics  with
a quantum efficiency in the Ce:LaBr$_3$ spectrum peak ($\sim$~380~nm) of
 $\approx 43\%$ \cite{baldazzi17}.
The PMT voltage divider has to deal with high instantaneous 
current density in presence of the pulsed muon beam and given the whole detector characteristics.
 A custom
active voltage divider was designed to ensure the best
performances of the system, both in timing and  spectroscopy, see reference
\cite{baldazzi17} for further details.

\subsection{Compact X-rays detectors with SiPM array readout}
\label{sec:SiPM}
Detectors more compact than
Ce:LaBr3 crystals with PMT readout are needed to equip the remaining regions, as the ones under the target vessel,
otherwise difficult to be instrumented 
 due to lack of space.
The choice was to use 1/2" crystals with a SiPM array readout, 
where the detector volume is only slightly bigger than the crystal
dimensions. 
In the 2016 run, two 1/2" Ce:GAAG and four 1/2" Pr:LuAG detectors were used,
for more details see reference \cite{bonesini15}..
In the 2017 and following runs, 
they were replaced by eight
 1/2" Ce:LaBr$_3$. 
The 1/2" crystals   are read by
  a $4 \times 4$ array
made of $3 \times 3$
mm$^2$ SiPM.
The detector holder was realized with a 3D printer. 
Operating voltages
 are set
according to manufacturer's specifications.
Most of our results were obtained with Hamamatsu S13361 arrays, based on a TSV
(``Through Silicon Via'') technology.
For Pr:LuAg with emission in the
near UV (NUV) and Ce:LaBr$_3$ crystals
SiPM arrays with a Silicone window  were
 used to increase
response at NUV wavelengths.
The output from each pixel of the $4 \times 4 $
SiPM array is summed up on
a custom PCB.
For more details see reference \cite{bonesini17a}. 

To correct for the temperature gain drift of the SiPM arrays,  
an online correction will be  implemented via a Nuclear
Instruments NIPM12 digital controlled power supply, with a temperature feedback
provided by a onboard temperature sensor. 

\subsection{The HPGe X-ray detectors }
In all the runs of the FAMU experiment, four HPGe detectors complemented
the Ce:LaBr$_3$ crystals for characteristic X-rays detection. Their main
aim, covering only a small fraction of the solid angle, was to provide 
a high precision  inter-calibration for the whole detection system, due
to the better energy resolution of HPGe detectors as compared to Ce:LaBr$_3$ 
ones.\footnote{As an example, at 122~keV an Ortec  GLP (GEM-S) detector has a 
nominal energy resolution of 0.4 $\%$ ($0.7 \%$)  FWHM to be compared with
$8 \%$ from a typical Ce:LaBr$_3$ detector.}
They are also useful to identify contaminations in the gas and background 
sources.

The four HPGe detectors were two Ortec GEM-S, one Ortec GLP and one Ortec GMX 
detector. 
The GEM-S detectors have a semi-planar geometry, p-type with diameter~$\times$~length 
30~mm~$\times$~20~mm and a 0.9~mm carbon window. The GLP detector has  a planar
geometry, n-type with diameter~$\times$~length 16~mm~$\times$~10~mm and a 0.127~mm 
beryllium window. The Ortec GMX is instead a coaxial n-type detector with 
diameter~$\times$~length
54.8~mm~$\times$~49.8~mm and a 0.127~mm beryllium window. 

HPGe detectors signals shaped through a preamplifier and a shaper
were sent, after a splitter, both to a CAEN V1724 100~MHz FADC and a Ortec 
MCB Multichannel Analyzer for a fast online analysis, via the MAESTRO 
software~\cite{maestro}.
For all the  HPGe detectors  Ortec~672 spectroscopic
amplifiers were used. In addition, for one  GEM-S HPGe detector also a fast
Ortec~579 shaper was used. The shaping time was 2~$\mu$s and 200~ns, respectively
for the Ortec~672 and the Ortec~579 modules.   
Studies are under way to develop a faster pre-amplifier, with a
$\leq100$~ns risetime, to be compared with a standard of $\sim200$~ns.

\subsection{Beam momentum tuning and detectors positioning}
The optimal disposition of the X-rays detectors around the target 
depends on the distribution
of the muon stops. X-rays coming from inert material,
especially during the arrival of the muon spill, must be minimized
while maximizing the X-rays produced in the gaseous target. The muon
stop in the apparatus depends on the amount and types of material along
the beam line and on the momentum and spatial spread of the beam.

A Monte Carlo simulation has been used to determine the best
configuration. 
The GEANT4 simulation toolkit~\cite{geant4}
(version 4.10.01) has been used in conjunction with the Generic GEANT4
Simulation software~\footnote{GGS v. 2.1, developed as a fork from the official
simulation setup written for the GAMMA-400 experiment by the Trieste
WiZard group and available via git: ``git 
clone git://wizard.fi.infn.it/GGSSoftware.git''.}. The software provides a set of tools for signal
hit readout and the possibility of a runtime plug-in detector geometry
implementation using external shared libraries.

The FAMU GEANT4 simulation geometry reproduces the lead collimator, the target, the
cryogenic vessel and detectors with their ancillary components, as insulating
multi-layer foils, supports and inert materials. 
The simulation takes into account the
geometrical physical dimensions and the materials used to
construct the target. 
\begin{figure}[!htb]
\centering
\includegraphics[width=0.60\textwidth]{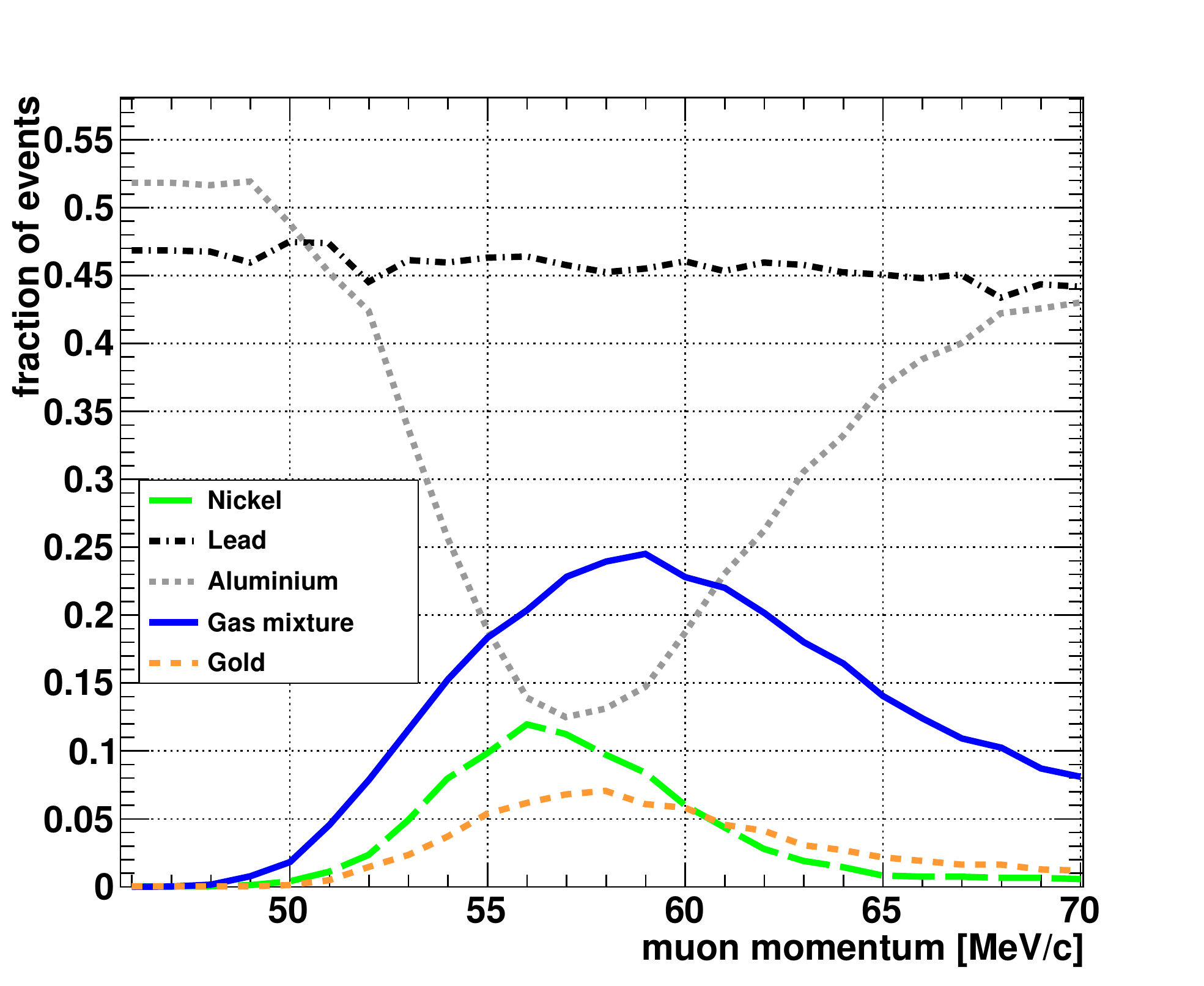}
\caption{GEANT4 simulation of the muon stopping in the FAMU apparatus.  
The figure  shows the fraction of events stopping in the gas
(blue solid line), in the Aluminium walls (gray dotted line), in the Lead
collimator (black dotted-solid line), in the coating (Gold yellow dotted line,
 Nickel green
line) as function of muon beam momentum.}
\label{figStop}       
\end{figure} 

The muon beam was simulated according
to~\cite{ishida03}, i.e. with a circular shape of
4~cm diameter, divergence of 60~mrad at the exit of the beam pipe and
a momentum spread  $\sigma_{p}/p \approx 4\%$.

By varying the beam momentum in the simulation, it was possible to
study the muon stop in the apparatus materials. Figure~\ref{figStop}
shows the resulting fraction of events stopping in different materials as a function of the beam momentum. 
It may be noticed how the
maximum muon stop in the gas almost matches the minimum stop in the Aluminium
at about~57 MeV/c beam momentum. The muon stop in heavier elements, as
Gold and Nickel, does not interfere with the transfer rate measurement
since the muon capture by the nuclei happens faster than the
thermalization and delayed phase of the muonic Hydrogen under study.

\begin{figure}[!htb]
\centering
\includegraphics[width=0.79\textwidth]{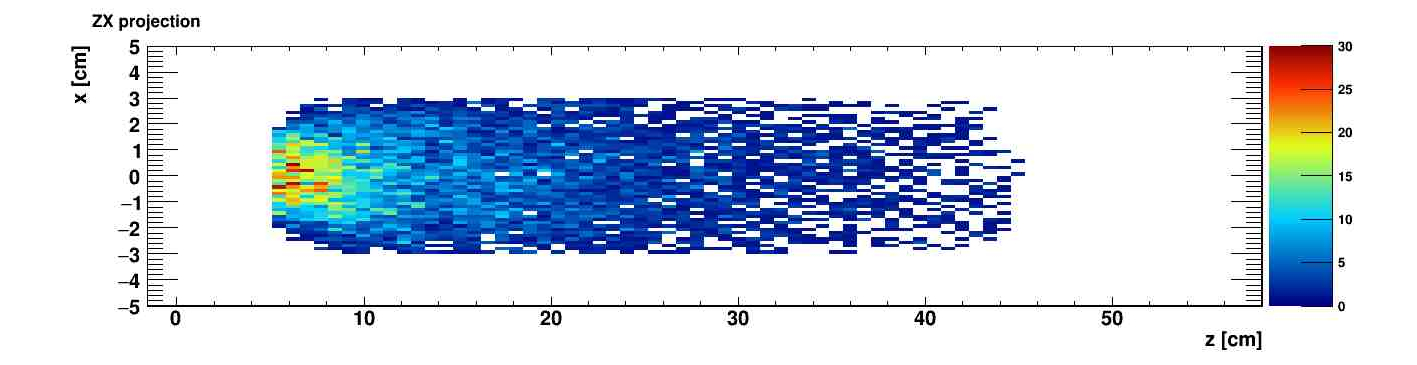}
\includegraphics[width=0.79\textwidth]{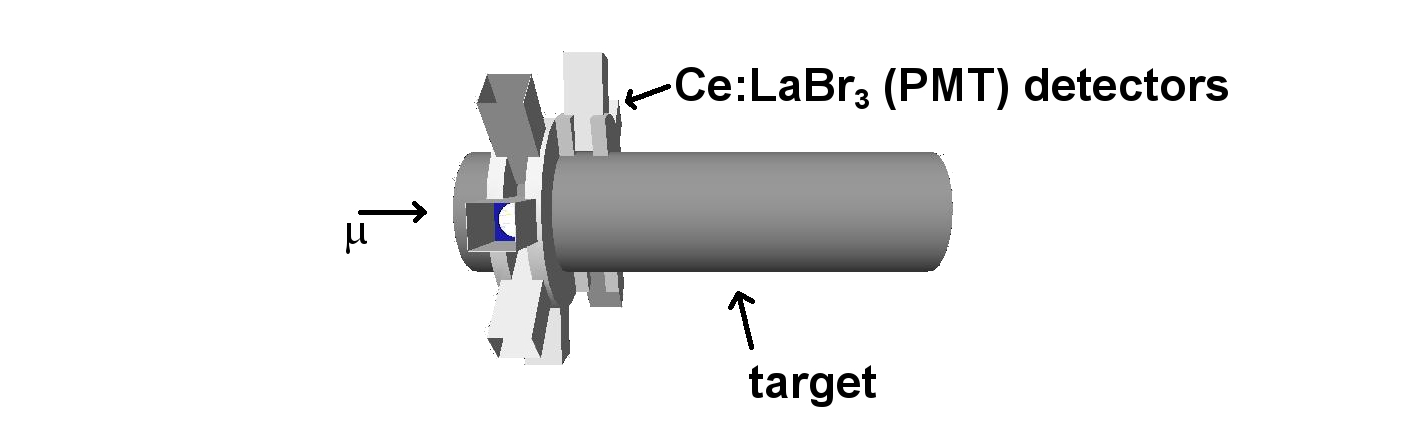}
\includegraphics[width=0.79\textwidth]{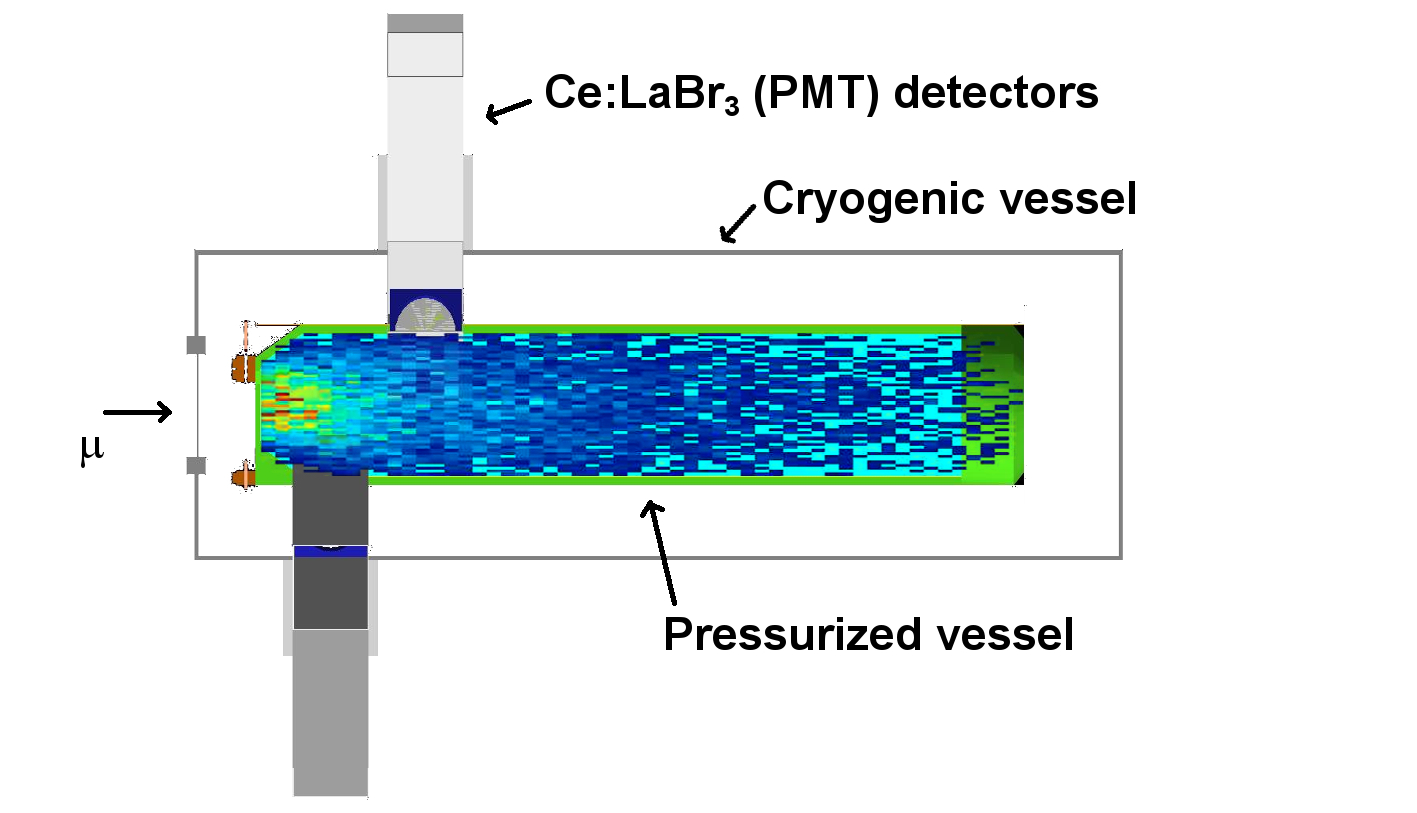}
\caption{Top panel: distribution of 57 MeV/c muons stopping in the
  target. Middle panel: drawing from the GEANT4 simulation of the
  target and the Ce:LaBr$_3$ crystals read by PMT. Bottom panel:
  section of the previous drawing with the muon stop distribution overimposed.}
\label{figStopDist}       
\end{figure}

Once the muon beam momentum has been fixed, it is possible to study
the spatial distribution of muon stop in the gas as shown in
figure~\ref{figStopDist}, top panel. The figure shows the ZX
projection of stopping muons --- the Z-axis being along
the beamline. 
Most of the muons stop in the front part of the
target, even if a fraction of them reach the rear, filling the whole target.

The fast Ce:LaBr$_3$ crystals read by PMTs were placed close to the
target in the region of maximum stop of muons in the gas. In 2016 data taking 
they were arranged in a single crown, while in the following periods they were
placed inside two half circular crown supports, as shown in
figure~\ref{figStopDist} middle panel. Bottom panel shows the
distribution of muon stops inside the simulated target.

\begin{figure}[!htb]
\centering
\includegraphics[width=0.41\textwidth]{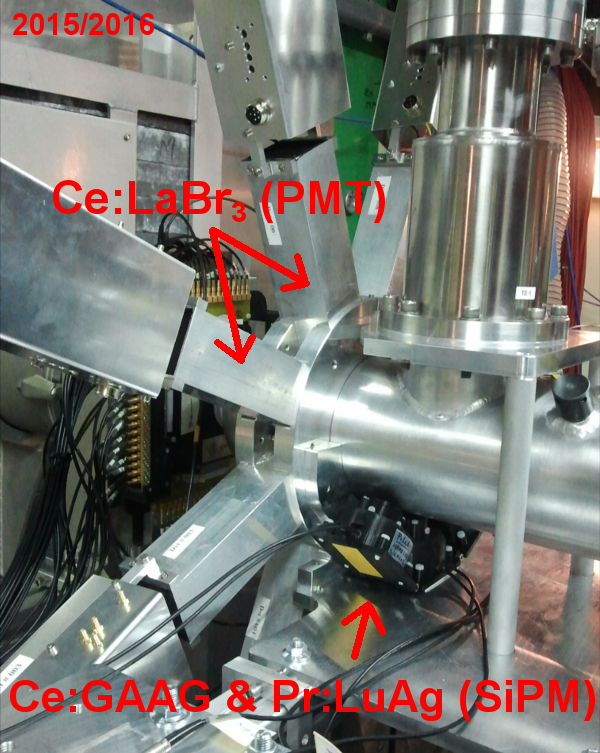}
\includegraphics[width=0.41\textwidth]{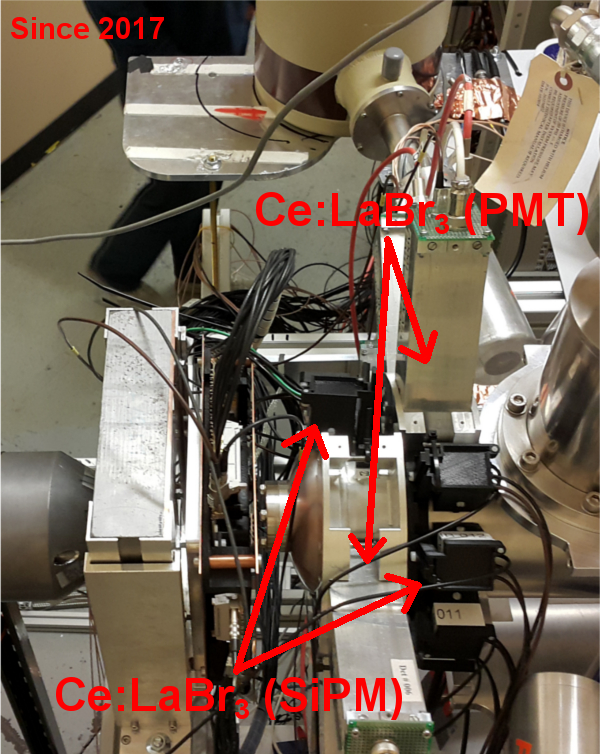}
\caption{Left panel: positioning of the T-shaped support structure
  with crystals read by SiPM (2015 and 2016 data taking). Right panel:
  positioning of the Ce:LaBr$_3$ crystal detectors since 2017.}
\label{mibpos}       
\end{figure}

Other type of crystals read by SiPMT, in 2015 and 2016, were placed,  thanks to
their compactness,  under the target
inside a T-shaped 3D-printed support, figure~\ref{mibpos} left panel. 
Since 2017, the Ce:LaBr$_3$ crystals read by SiPMTs 
were instead positioned in the front part
of the target using two half circular crown supports which completed the ones of Ce:LaBr$_3$ read by PMTs, as shown in
figure~\ref{mibpos} on the right.

HPGes detectors, due to their longer shaping time, are more sensitive
to X-rays pile-up. For this reason their position was farther away
from the target respect to the other detectors, usually pointing to
the rear part of the apparatus.
\begin{figure}
\centering
\includegraphics[width=.80\textwidth]{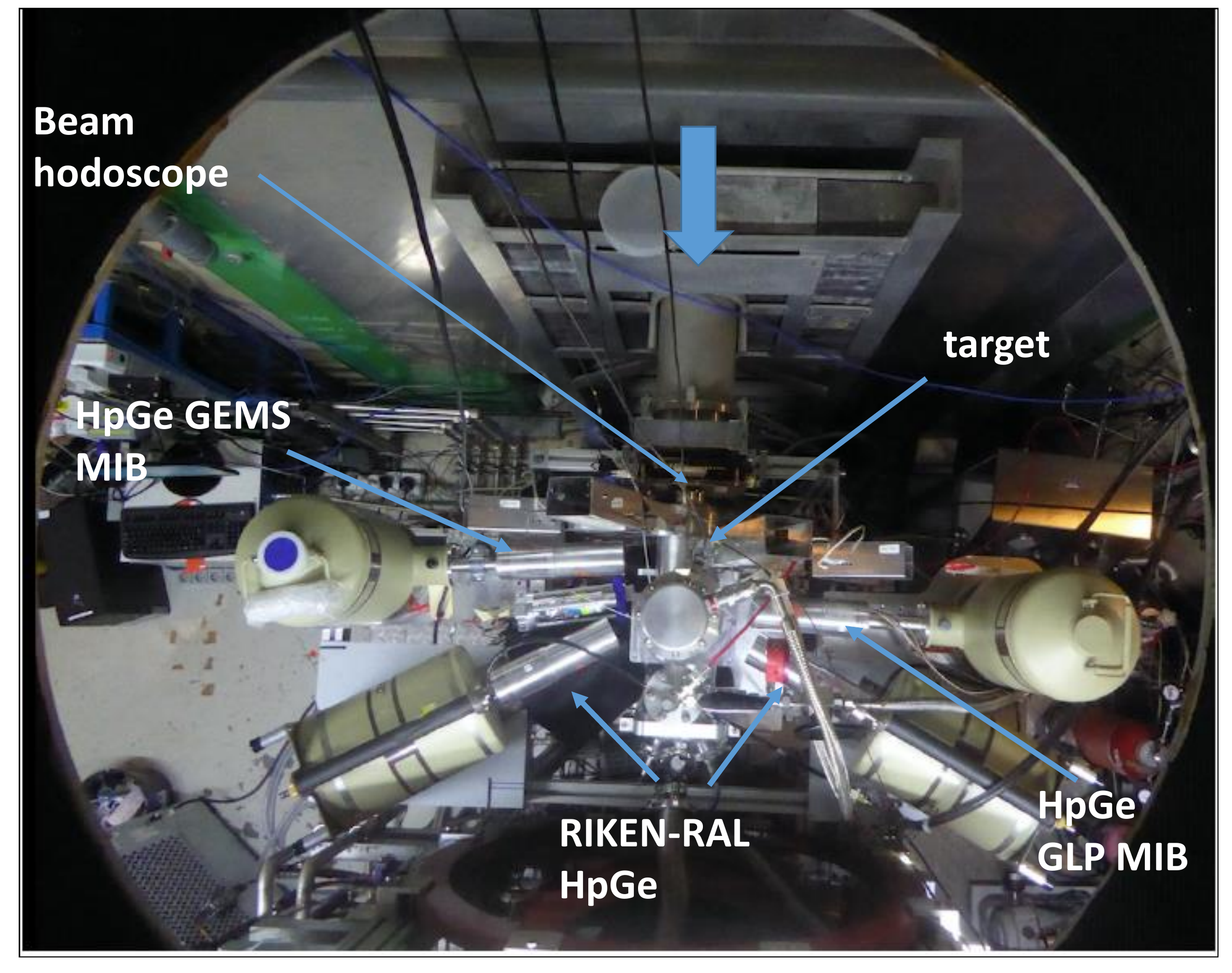}	
\caption{Picture of the FAMU layout seen from the top. The position 
of the HPGe detectors  and the 1~mm pitch beam hodoscope are shown.}
\label{fig:hpge}
\end{figure}
Figure \ref{fig:hpge} shows the position of the HPGe detectors
during the 2016 acquisition. Also shown is the hodoscope placed in
front of the target.


\section{Raw data handling}
The data processing from online to offline analysis is outlined in the 
following.

Signals (64 in all) from each hodoscope detector are digitized by two CAEN 
V1742 digitizer in VME standard, while signals from the crystal detectors 
(eight  with PMT readout and eight with SiPM array readout) are
digitized by V1730 or DT5730 digitizers.
The signals from the four HPGe are instead 
digitized by a CAEN 100~MHz V1724 digitizer. 
The V1742 modules are  read via a CAEN V2718 VME-PCI
interface. Instead 
the V1730, DT5730 and V1724 digitizers are read directly with an 
optical fiber via a CAEN A3828 PCI card, to have a faster transfer rate. 
The beamline gives an adjustable pre-trigger signal, typically 500~ns before the center of the
time at which the first muon spills reaches the target. Muons are bunched in 
two spills, with 70~ns width, at 320 ns distance. Thus the chosen windows of 1 
$\mu$s for hodoscopes, 10 $\mu$s for Ce:LaBr$_{3}$ detectors 
and 
20 $\mu$s for HPGe detectors were enough for containing the full signal 
waveform. The waveforms were sampled every 2 ns for the hodoscopes and the
crystal detectors and every 10 ns for the HPGe detectors. 
From signal waveforms the most
 complete set of information may be deduced, such as
peak  position, timing, etc. 
Runs of 10000 events were recorded on disk by a custom made data 
acquisition software
and stored as PAW ntuples. These ntuples are converted later to ROOT
 files for offline analysis. 

Further details on the online data acquisition system 
are shown in reference \cite{soldani}.
 
\subsection{Structure of offline data processing } 
\label{sec:offline}
Data processing sofware is written in C++ language,
and  makes use of the ROOT toolkit~\cite{root}. Acquired raw data are
in PAW ntuples format. They are converted in ROOTples containing the
 packed detectors waveforms.

Raw data (PAW and ROOT files) are automatically transferred from  the local DAQ storage disk to a remote repository 
via a Grid-FTP protocol. 
The main data processing program reads the raw files and retrieves the
observables needed in the analysis. 
\begin{figure}[!htb]
\centering
\includegraphics[width=0.9\textwidth]{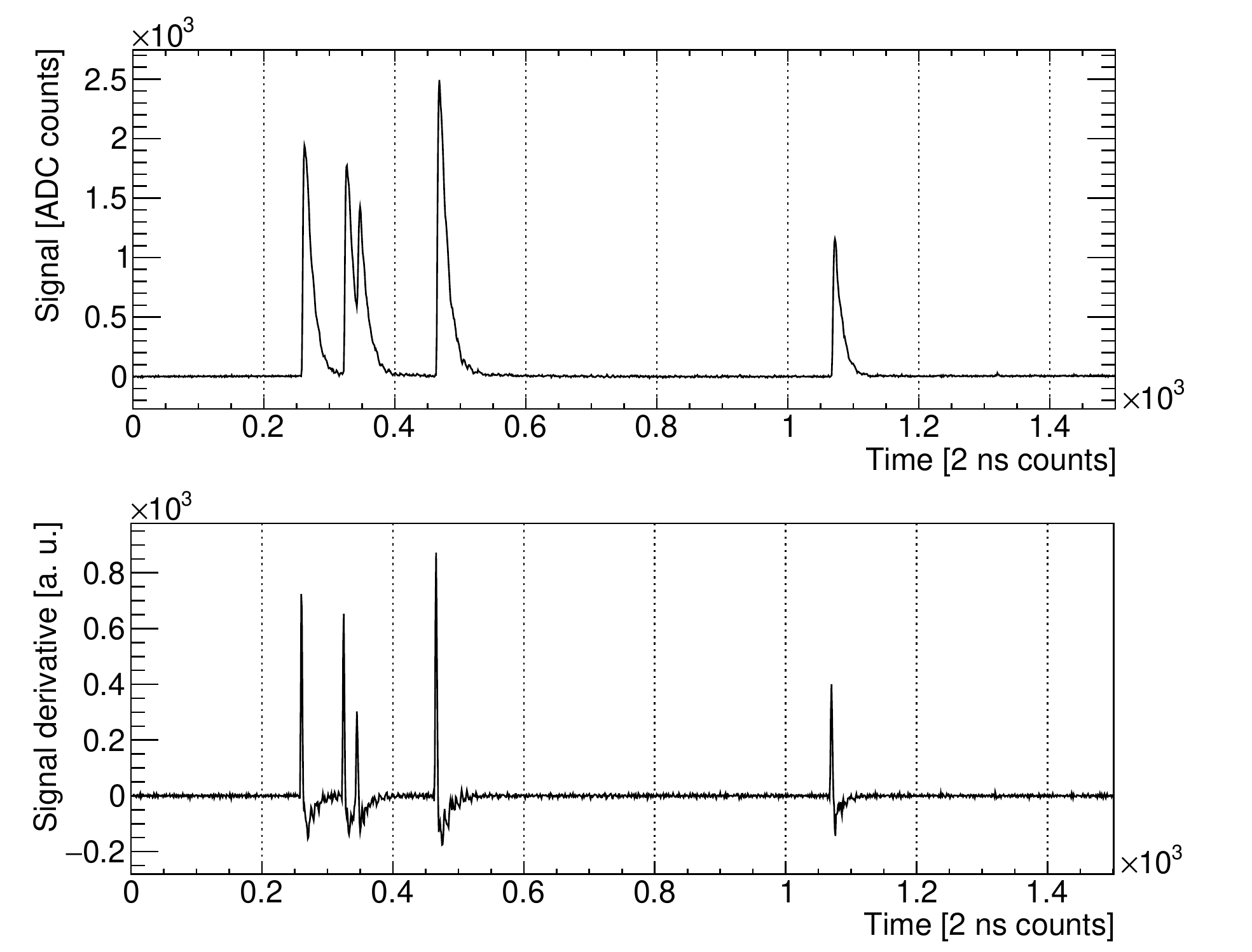}
\caption{Top panel: unpacked waveforms for a typical Ce:LaBr$_3$ detector channel.
Bottom panel: computed signal derivative. To each wavefront corresponds
a large spike in the derivative. One  count corresponds to two nanoseconds.}
\label{figSig}       
\end{figure} 

The data processing software has two main operating modes. The
first one is very fast (the processing time is smaller than acquisition time) 
and requires very little CPU power. This mode (``quick-look'') 
is used to monitor
 the data flow during the acquisition procedure
and it is used to produce data quality plots. The second mode gives
more accurate results for the data analysis but it is significantly slower (the processing
time is about ten times greater than the acquisition time), since fitting
algorithms are involved.

The data processing program treats the output from the various
detectors in different ways that depend on the waveform
shape and detector characteristics.

Hodoscopes are used to study the muon beam shape and intensity. 
The processing program
has a dedicate routine to process their waveforms. Both in the
``quick-look'' and in the ``full analysis'' modes, the
software reads the raw ROOT input file, measures the total energy deposit, above a fixed threshold,
of each of the fiber  and saves the data in a new ROOT file.

X-rays detectors (HpGe, Ce:LaBr$_3$ and other crystals) have most 
of the core routines in common.
For all the detectors and in both the processing modes, the program flow
is the following:
\begin{itemize}
\item for each detector the waveforms are unpacked event by event,
  as shown in  the top panel of figure~\ref{figSig};
\item a numerical derivative of the waveform is computed with a finite
  difference method using a five-point stencil in one dimension. In addition,
  resulting sharp edge due to signal saturation are smoothed by recognizing 
the flat top of the
  saturated waveforms. An example of signal derivative is shown in
  figure~\ref{figSig}, bottom panel. A fixed threshold is used to find the X-rays signals.
  The software identifies the starting point of the signal window when
  the derivative rises above the threshold. The approximate
  position of the maximum is given by the derivative zero
  crossing. The threshold value was set empirically in order to
  minimize the number of fake triggers, due to the intrinsic baseline
  noise, and maximize the number of detected X-rays. The determined
  threshold value on the derivative depends on the detector and, e.g. for 
Ce:LaBr3 read by PMTs,
  corresponds to about 50~keV X-rays signals.
\item the energy of each X-ray signal is reconstructed together with its
  arrival time;
\item the amplitude of each signal and its starting time is saved in
  the output data structure, together with other informations
  (number of the detector, and
  housekeeping informations as the target temperature and the
  absolute linux time);
\item the output data structure is saved in an output ROOT file.
\end{itemize}
The same program checks the data for signal saturation (mostly due to
the digitization electronics) and saves the information in the same
output file. The method to determine energy and starting time of each X-ray depends
on the detector and on the analysis mode. 

The fastest detectors are the Ce:LaBr$_3$ crystals read
by PMTs. They have been used to study the physics of transfer rate of the
muon from muonic Hydrogen to other elements  in the 2016 data taking.
The energy of each signal is reconstructed with two different
 algorithms, depending on the operating mode. 
The first one, used in the ``quick-look'' processing, determines the difference between the
  baseline and the maximum of the waveform in
  each identified signal window. This method is fast and works at its best with a low
  fluence of X-rays, but it is able to resolve with a low efficiency the overlapping signals 
  (pile-up). 
Instead  in the ``full analysis mode'', each signal is fitted with a function
  representing the signal shape. In case of pile-up a sum of signal-shape functions is used in order
  to measure correctly the energy of each X-ray.
\begin{figure}[!htb]
\centering
\includegraphics[width=0.6\textwidth]{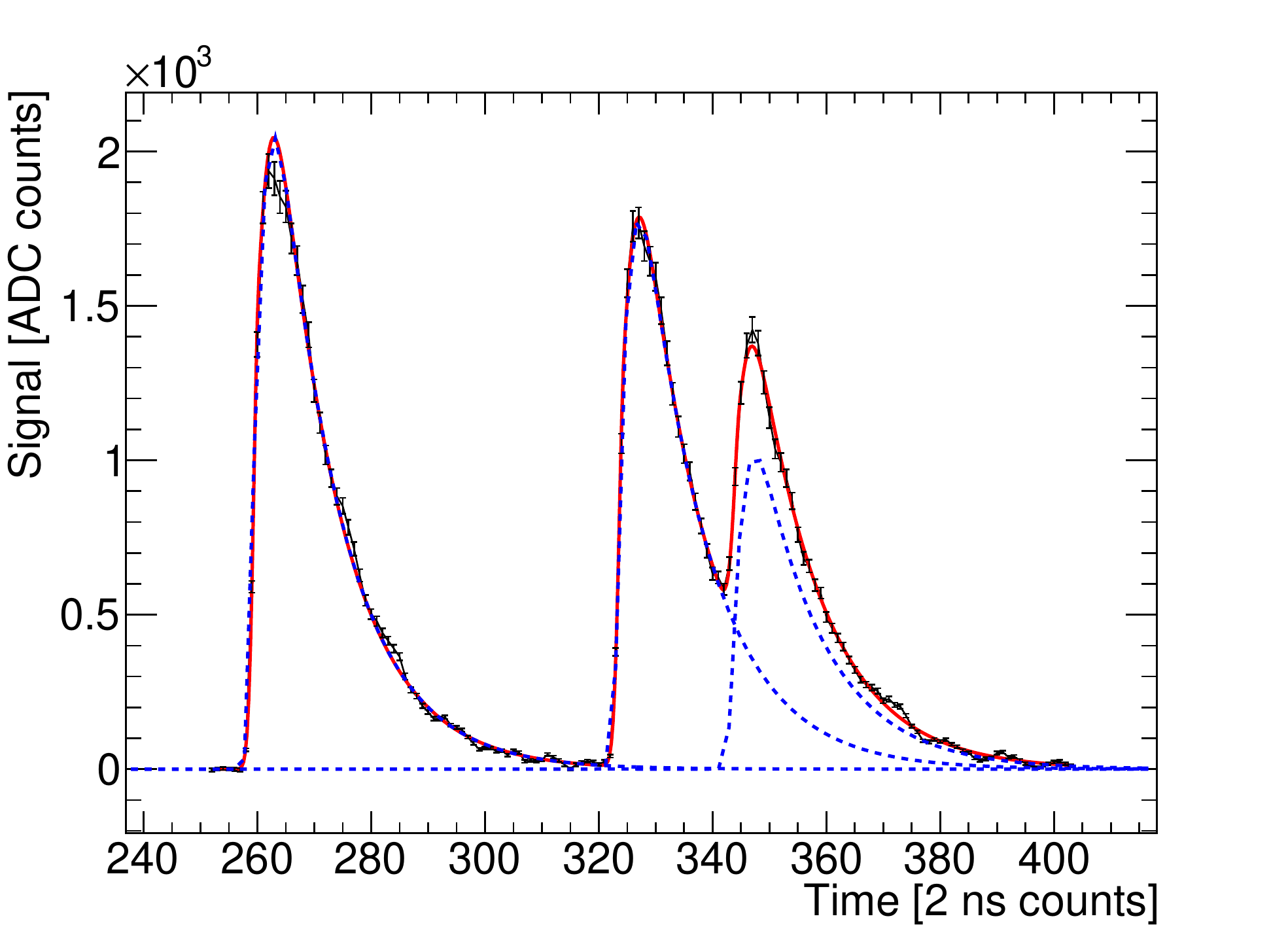}
\caption{Example of multiple overlapping signals. Black points represent the data waveform, red line is a fit of the data using three signal shapes shown as blue lines.}
\label{figSigFit}
\end{figure} 
  As an example, in figure~\ref{figSigFit} the red line represents the total fit
  function, the blue lines the single components due to each X-ray,
  while the black points are the data to be fitted. The normalized
  $\chi^2$ of the fit, the degrees of freedom and exit fit status are
  stored together with the data. 
This pile-up
reduction by the multiple fitting algorithm is the most compelling task of the data processing. A fine tuning of input
parameters and the excellent stability of the detectors waveforms
allowed a reliable reconstruction of piled up events. The
software was tested using simulated waveforms (with included
simulated electronic noise). Placing constraints on the resulting
$\chi^2$ of the fit and on the distance between two neighbour signals,
an efficiency greater than 90\% was obtained for X-rays detected in the delayed phase, where the transfer
rate is going to be measured. Efficiency drops to about 50\% when
the time distance between two X-rays signals is shorter than about
30~ns. In this situation, which is present only during the prompt
arrival of muons in the target, only the first signal energy is correctly
reconstructed. The ``perfect pile-up'', i.e. the situation in which the software
is not able to distinguish two signals, starts at a separation of less
than 15 ns between signals. A Monte Carlo simulation proved that this
phenomenon is negligible in the delayed phase.

The HPGe detectors amplified signals are usually too slow to
detect many X-rays during the same trigger. Hence, the processing
software for both the ``quick-look'' and the ``full analysis'' modes scan the
whole waveform and save the maximum of the signal amplitude and its time,
one X-ray per each trigger (the information about signal time windows
is not used in this case).

HPGe pre-amplified and fast-amplified signals and other fast
X-ray detectors signals (Ce:LaBr$_3$ and
other type of crystals read by SiPMT) are processed using the same
procedure. The ``quick-look'' mode uses the same maximum minus mininum
algorithm described for germanium amplified signals. The ``full analysis''
mode, instead, look for signal maximum in each signal window, the
same algorithm described for the quick-look mode of Ce:LaBr$_3$ read by PMTs.

Finally, data for each detector are calibrated offline in order to convert from 
arbitrary digitizer units to keV. This is usually done by processing
the source calibration data taking and by knowing the characteristic X-rays
nominal peak positions
of the source. Calibration parameters are saved in a MySQL database and
used in the following data analysis.

\section{Experimental operations and performances}
The operations of the various elements of the FAMU experiment at RIKEN-RAL
will be described in detail in the following. 

\subsection{Target operations }
\label{target_op}
A first set of measurements on the RIKEN-RAL muon beam, with the FAMU high 
pressure gas Hydrogen target and detectors  allowed us to successfully 
validate the target system.

The target filling procedure requires first a turbo-molecular
vacuum pump to clean the gas target. 
The target at 300~K temperature is evacuated until the pressure is measured to be as low as
$5\times 10^{-6}$~bar. Three cycles of gas filling (at 10~bar)
and flushing with pure Hydrogen are  then performed. After
having re-established
 the vacuum with the turbo-molecular pump, the gas mixture needed for
the measurement is injected in the target.

At any injection the measured temperature increases up
to 10 K, due to the heating of the gas when crossing
valves and tubes. This effect was studied carefully in 2017 by
measuring the pressure and temperature of the target during and after
the filling. For example, the target filled at 40 bar starting at 300~K increases
its temperature up to about 310~K; when the temperature is brought
back to 300~K the measured pressure is of 39~bar. This means that a
drop in the pressure of about 2.5\% is observed respect to the nominal
filling pressure. In 2016 data taking, the pressure was measured only
at filling time, so the related data analysis has to include the
2.5\% correction in the pressure measurement.

Once filled and thermalized, the target is used for the measurements
at different temperatures. 
\begin{figure}
\vskip -2cm
\centering
\includegraphics[width=0.9\textwidth]{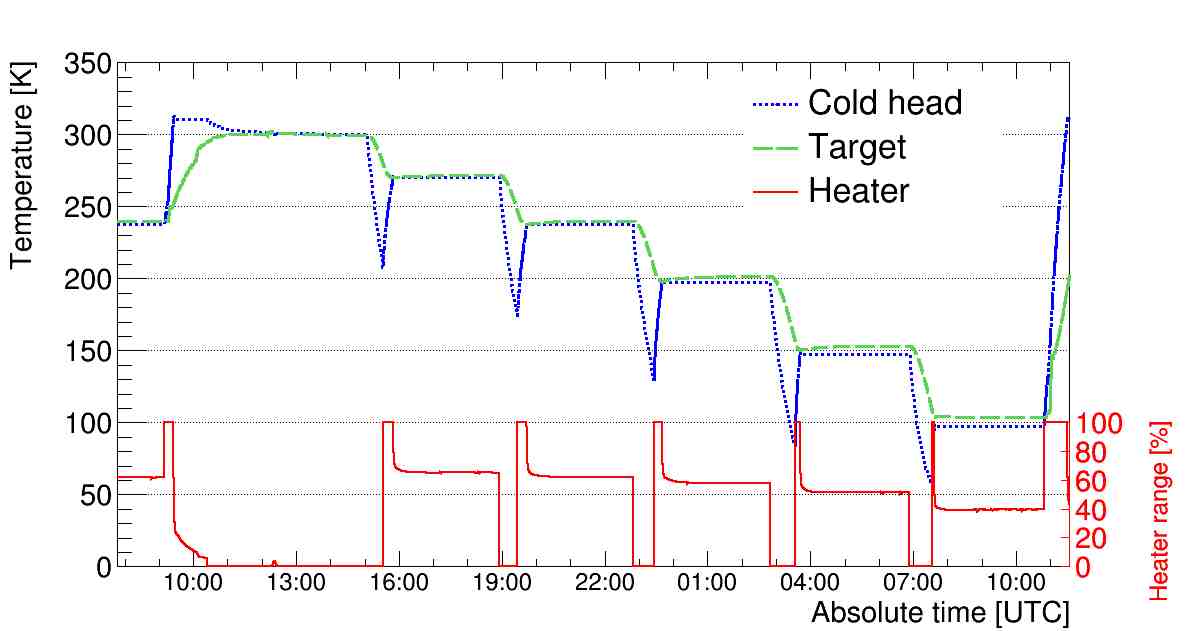}
\caption{Example of a temperature cycle performed on the beam with the
target in 2016. Blue line cold head sensor, green line target sensor, magenta
line heater power percentage.}
\label{fig:oper}
\end{figure}
An example of temperature cycle is shown in 
figure~\ref{fig:oper}, during the 
February 2016 data taking with $H_2+O_2(0.3\%)$
gas mixture. After the filling, at
10:00 a.m., 
three hours of data are acquired. The blue line represents the cold head
temperature measurement. It varies very fast depending on the power
induced by the thermo-resistance heater. The percentage of power is
shown as a magenta line in the same figure. The temperature of the
target, green line, varies much more
smoothly, since the heat is brought to the target with copper
connections. As described in section~\ref{sec:targetsystem}, to
make the transition between different temperatures faster, the cold
head was brought to a lower temperature than the target one and only
when the target temperature was approaching the desired value the cold
head was heated again. This can be clearly seen in the figure at about
16:00, where the cold head temperature goes down to
about 200~K while the target temperature is changes from 300 to 273~K. 
Six steps of three hours each were taken at the target temperatures of
300, 273, 240, 200, 150, and 100~K. Once the cycle ends, the target is
heated to 300~K to clean up and flushing and to start a new cycle
with a different gas mixture.

\subsection{Beam characterisation with the 1~mm pitch hodoscope}
\begin{figure}[htbp]
\centering 
\includegraphics[width=.70\textwidth]{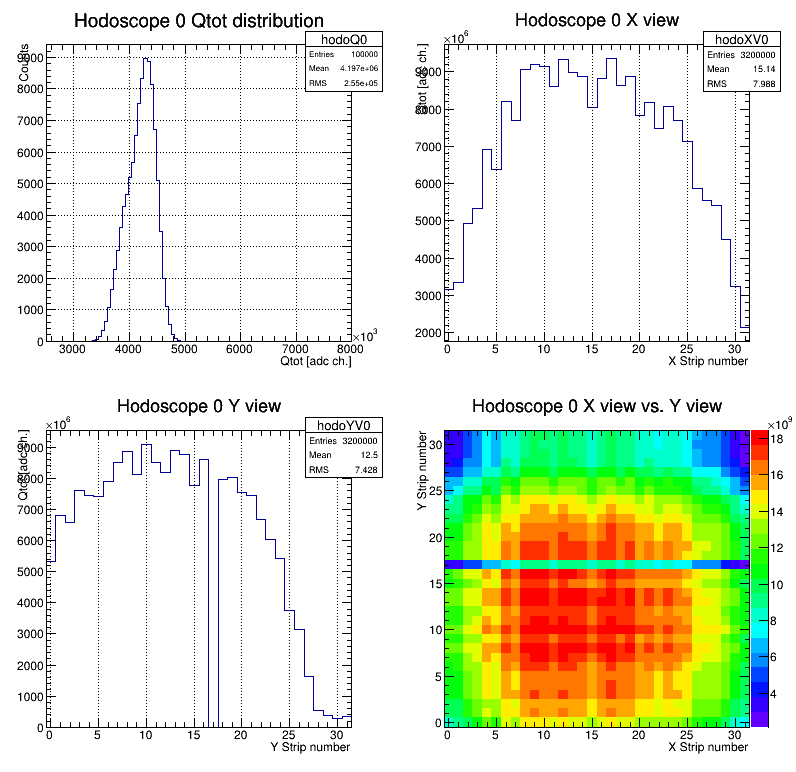}
\caption{X/Y beam profile at RIKEN-RAL for a 57~MeV/c run.The accumulated total
         charge is also shown. }
\label{fig:res2}
\end{figure}

The  
1 mm pitch beam hodoscope has been used at RIKEN-RAL to control and 
optimize the
beam steering inside the target. The signal waveform for each channel
was integrated, after subtracting the baseline, to provide informations for
an X/Y occupancy plot. 
The X/Y beam profile for a typical run is shown  in figure \ref{fig:res2}.
The total charge collected on the beam hodoscope ($Q_{tot}$) is also shown and may be used
to monitor the beam intensity. A channel of the hodoscope was dead in this 
particular run, taken in 2016. This was due to a broken signal cable.

\begin{figure}[htbp]
\vskip -1cm
\centering 
\includegraphics[width=.80\textwidth]{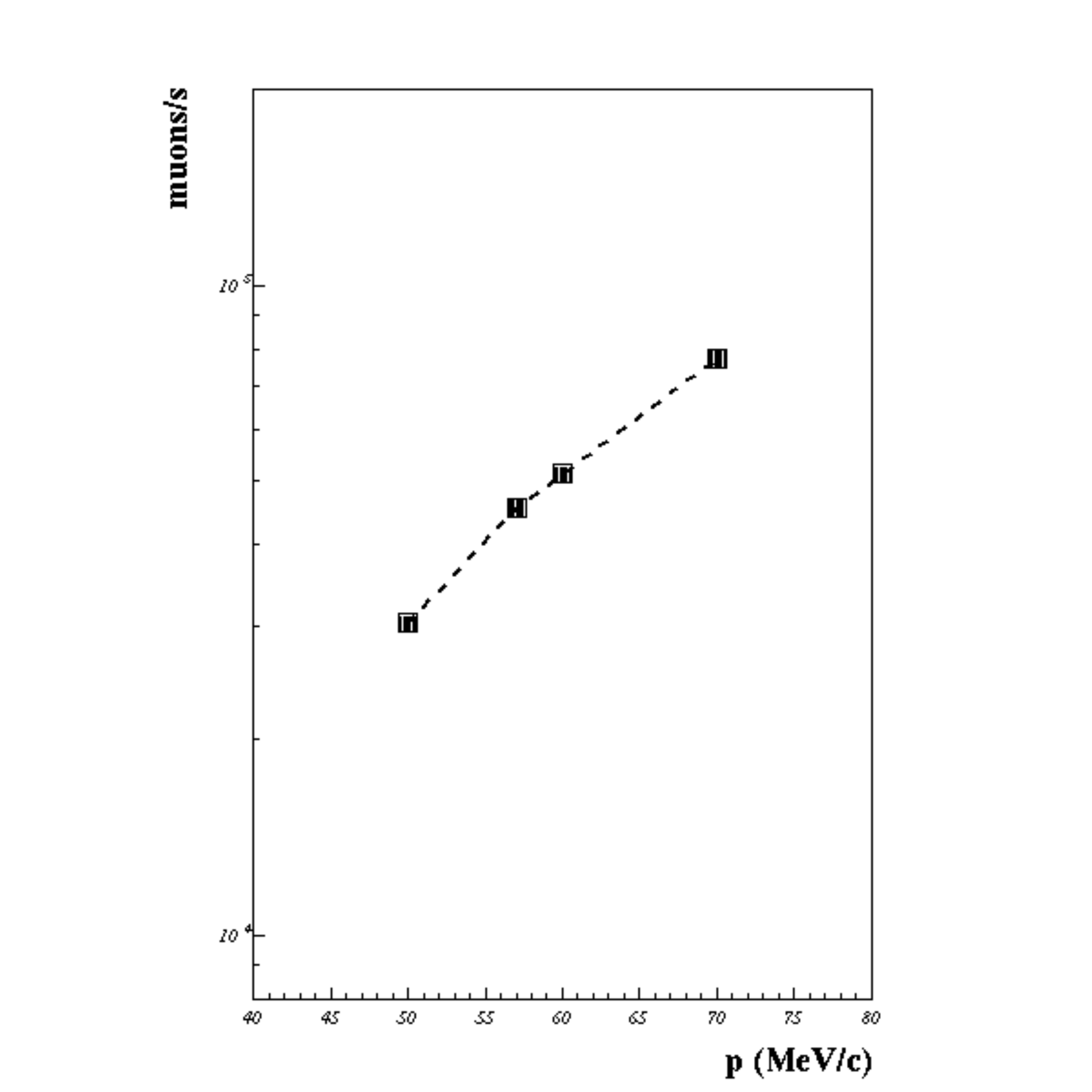}
\caption{Muons per second at the target window, as measured in a scan during 
2018 data taking. }
\label{fig:scan}
\end{figure}

Making use of laboratory data taken with impinging cosmic rays, and using
tabulated dE/dx data from PDG~\cite{pdg}, it was possible to estimate from
the total charge collected ($Q_{tot}$) the number of beam muons. The spill rate (nominally 40~Hz) was estimated from
data itself and was $\sim$30~Hz, as only three out of four spills 
reached the target in these runs. 
An example of the muon rate per second 
 is shown in figure \ref{fig:scan} for a momentum scan taken in 2018.
Results are compatible with previous results, reported in \cite{matsuzaki01}.

\subsection{Detection of characteristic X-rays with Ce:LaBr$_3$ with PMT 
readout }
\begin{figure}
        \centering
        \includegraphics[width=1\textwidth]{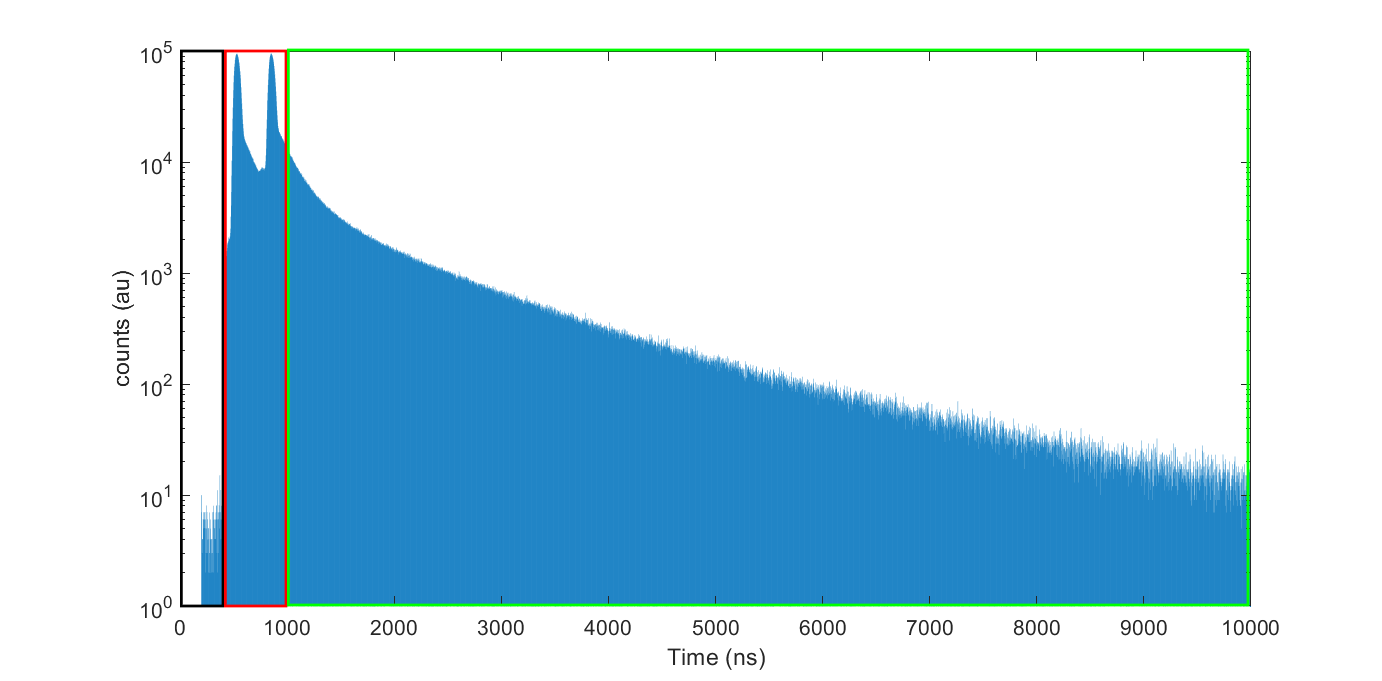}
        \caption{Time spectra from Ce:LaBr$_3$ with PMT readout. 
Three regions are indicated: pre-trigger in black, prompt phase in red and delayed region in green.}
        \label{fig:PMT_time_spectra}
\end{figure}

Due to the high events rate, the minimum pile-up condition is preferred to the highest possible energy resolution condition.\\
The typical acquisition time window is 10~$\mu$s in which the  PMT 
output signal is digitized at 500~Ms/s (2~ns time step).
The whole window is divided in three zones (figure~\ref{fig:PMT_time_spectra}):
\begin{itemize}
        \item pre-trigger, a small low counting zone, right before the arrival
of muons  [0 ns--400~ns];
        \item prompt phase, where muons arrive from the accelerator 
          beam pipe and interacting with the target produce most 
          X-rays events [400~ns--1000~ns];
        \item delayed region, the region of interest for the FAMU experiment. 
All the muonic X-rays emitted by transfer phenomena are produced here [1000~ns--10~$\mu$s].
\end{itemize}

\begin{figure}
        \centering
        \includegraphics[width=.4\textwidth]{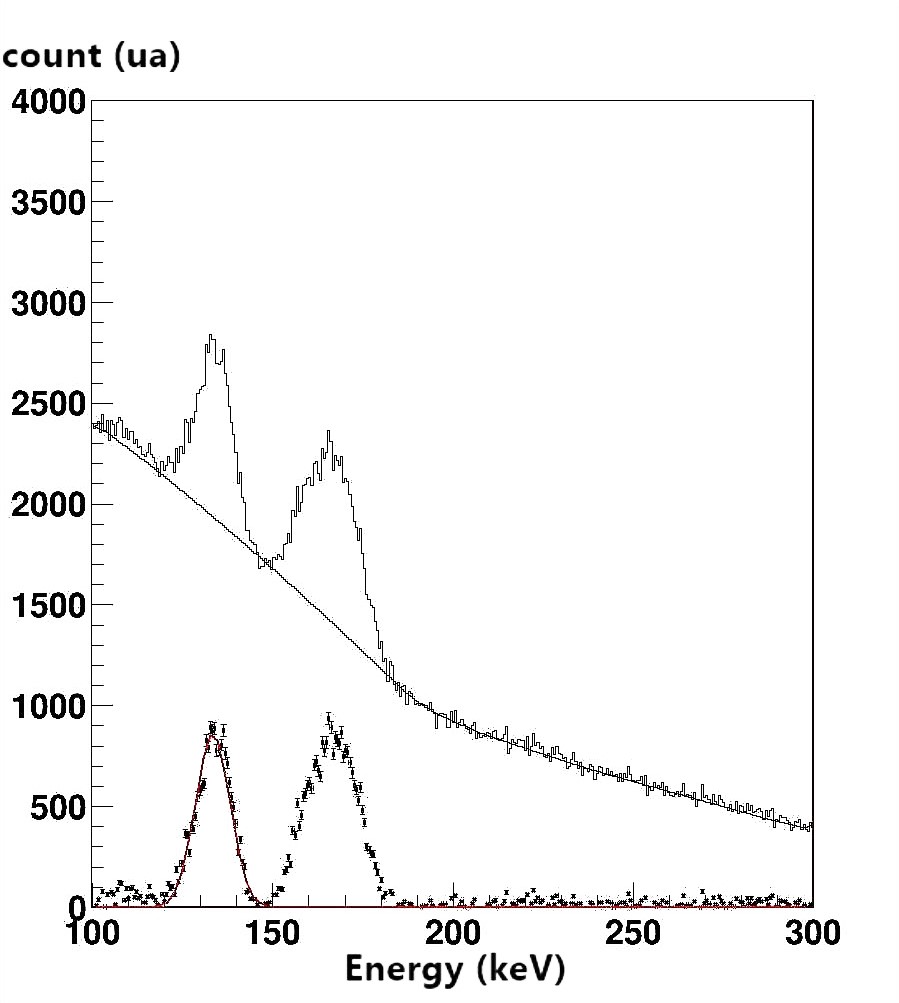}
        \caption{Measured delayed spectrum with a $H_2+O_2 (0,3 \%)$ gas mixture
in the target. Muonic Oxygen X-rays: 133~keV
K$_{\alpha}$, 158~keV K$_{\beta}$ and  167~keV K$_{\gamma}$ lines, with
background (top) and background subtracted (bottom).
The energy resolution for the 133~keV line is 8.5\%.}
\label{fig:133kev}
\end{figure}

\begin{figure}
\centering
\includegraphics[width=.9\textwidth]{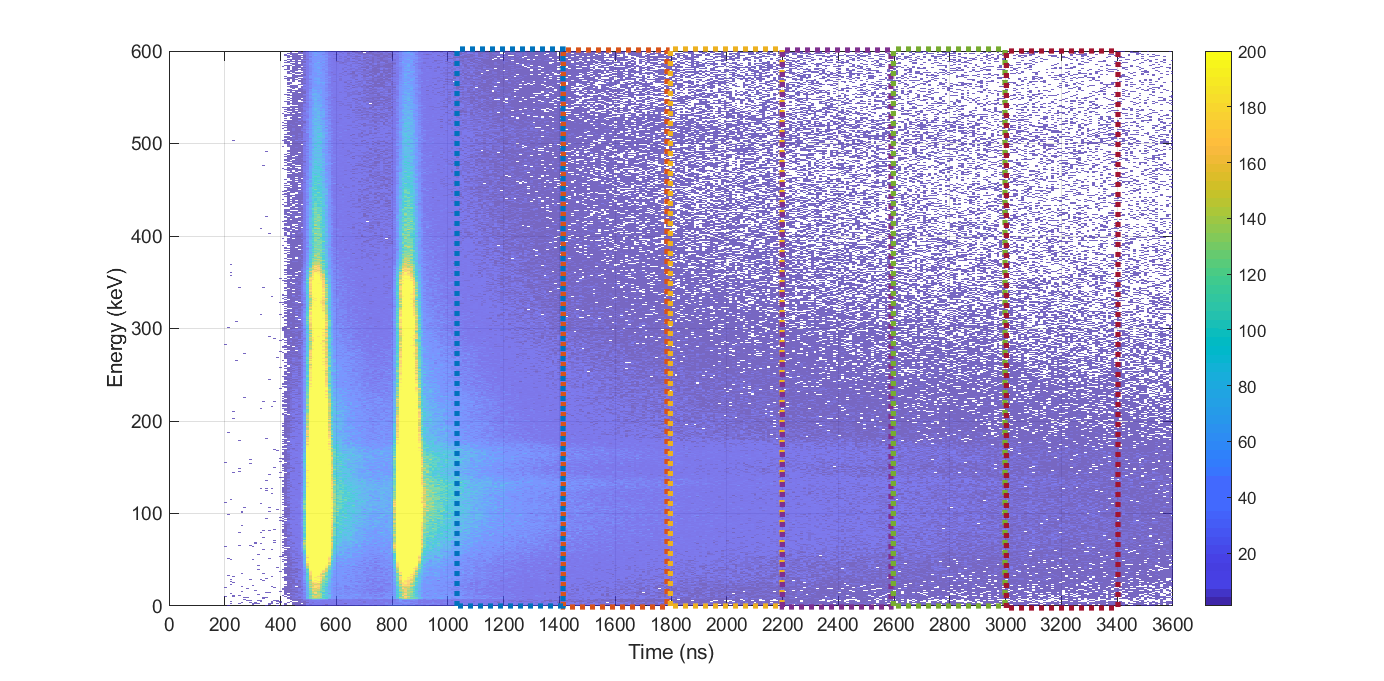}
\includegraphics[width=.5\textwidth]{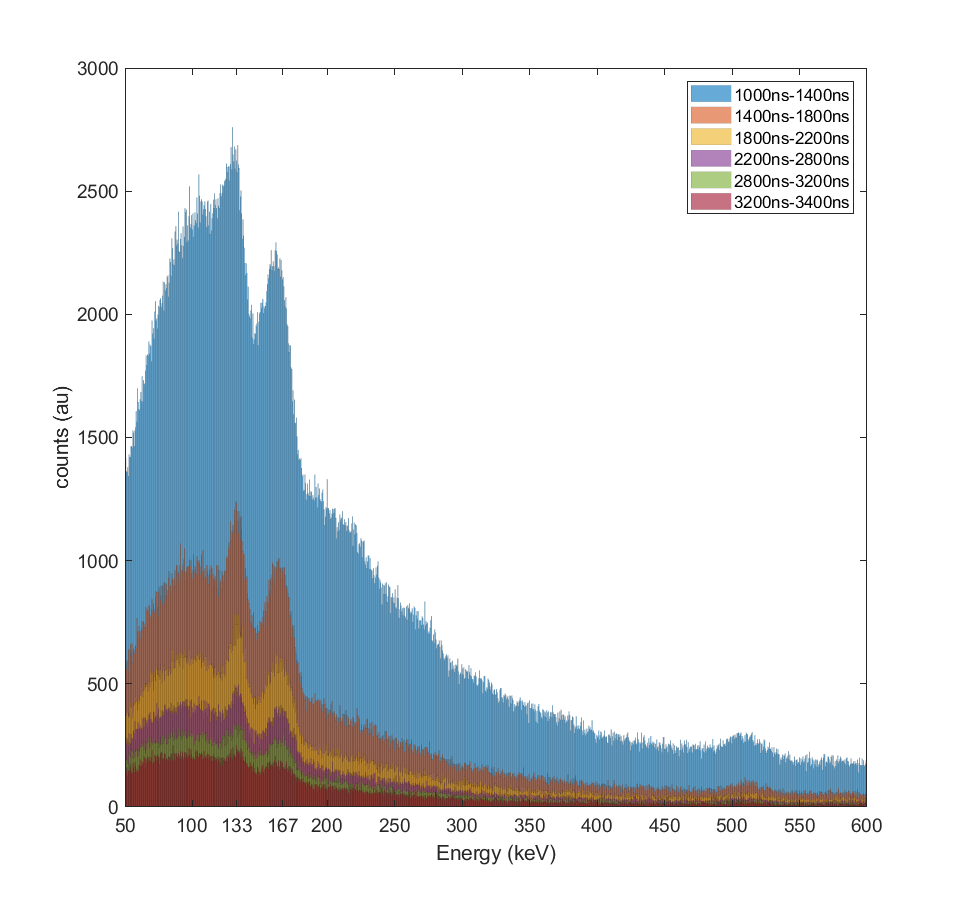}
\caption{Top panel: X-ray time evolution spectrum of the H$_2$--0.3\%O$_2$ mixture with a muons momentum of  P = 57~MeV/c.
Bottom panel: 400~ns integrated time cut spectra right after the prompt phase (from 1000~ns). All the time cut selections are shown in the top panel.}
\label{fig:3Dphys}
\end{figure}

At the muon beam arrival, 
an average number of 4 X-rays are detected in a $1~\mu$s time slot, and up to 
10 X-rays are detected in a 10~$\mu$s time window (see figure~\ref{figSig}).\\
The detector signal for a typical X-ray event has a peaking time 
of $\sim$12~ns and an exponential decay constant of
$\sim$16~ns. The PMT gain is set to fit the data acquisition dynamic range 
with maximum energy of about 1~MeV. A custom algorithm was developed 
to identify each single pulse in the full acquisition window, 
even in case of pile up (see section~\ref{sec:offline} for details). 
Energy and time of arrival for each X-ray event are stored. 
The algorithm is capable to resolve pile-up events down to 15~ns of peak 
separation. From 15~ns to 30~ns it has an efficiency of $\sim$ 50\%. 
It reaches a 90\% efficiency 
after 30~ns of separation. This allows to record 
the largest number of events (increased statistic) losing only a small fraction of events due to pile-up.
The priority given to the minimization of pile-up affects the energy resolution.
\begin{table}
 \caption{Obtained FWHM resolution of Ce:LaBr$_3$ crystals with PMT readout as reported in literature and measured from calibration runs with
\textsuperscript{57}Co and \textsuperscript{137}Cs sources.}                \centering
\smallskip       
         \begin{tabular}{ | l | c | c |}
                        \hline
                        Energy & FWHM from ref.  \cite{LABR2}
&measured FWHM  \\ \hline
                        122 keV & 7.4\%  & 8.8\%  \\ \hline
                        662 keV & 2.8\% & 3.5\%  \\ \hline

                \end{tabular}
\label{PMT_res}
\end{table}
It is slightly worse as respect to the achieved ones in literature
(see table~\ref{PMT_res} for further details).
To evaluate the detector performances during the data taking 
(in high instantaneous X-ray flux conditions) we measured the energy 
resolution of the 133~keV muon Oxygen $K_\alpha$ line acquired right after 
the prompt phase (1000~ns--10~$\mu$s).
Theoretically, energy resolution varies in respect with energy, 
following an inverse quadratic rule, so it is possible to extrapolate 
the expected resolution at any energy from  calibration data.
In particular, using the 8.8\% value as FWHM resolution for the 122 keV 
\textsuperscript{57}Co source (see table~\ref{PMT_res}) 
we expect 8.4\% at 133~keV. Experimentally we found 8.5\% 
(figure \ref{fig:133kev}) that is compatible within the total systematics uncertainties. This demonstrates the good detector performance in both high and low rate conditions.
The good detectors performances, combined with a proper 
analysis fitting algorithm have permitted to achieve both good spectral and 
timing performances with excellent detection efficiency.
 These performances fit well with the experimental requirements: 
study the X-rays spectrum time evolution from the prompt region to 
the delayed one.
The time evolution of the X-rays emission 
of the Hydrogen (99.7\%) Oxygen (0.3\%) mixture is shown in figure~\ref{fig:3Dphys}. 
It is possible to notice that the 133~keV $K_\alpha$ and 158~keV + 167~keV (respectively $K_\beta$ and $K_\gamma$) 
muonic Oxygen lines 
are emitted for few hundreds of nanosecond after the prompt ones.
 During the first phase of the FAMU experiment, these detectors allowed
the high precision studies on the transfer rate from muonic Hydrogen 
to Oxygen \cite{vacchi16,mocchiutti17}.

\subsection{Detection of characteristic X-rays with crystals with SiPM array readout }
Pr:LuAG and Ce:GAAG crystals read by SiPM arrays 
(see section~\ref{sec:SiPM} for details) were placed under the cryogenic target 
in the 2016 data taking
as additional X-rays 
detectors. They were initially calibrated in situ 
with a $^{57}$Co source and a
source that was an admixture of $^{241}$Am, $^{90}$Sr and $^{137}$Cs. 
As an example, at the $^{137}$Cs peak for a Ce:GAAG detector  the 
energy resolution (FWHM) was $\sim 6.9 \%$. 
Due to their poor positioning in the detector layout, in the 2016
data taking it was not possible to detect Oxygen X-lines from delayed events, 
even if the detectors were able to see the two beam pulses structure inside a spill 
(see reference \cite{alessandro2018} for further details).

In the 2017 run, the previous 1/2" crystals were  replaced by 1/2" 
Ce:LaBr$_3$ crystals, to improve the energy
resolution. The general detector's structure was left unchanged. In the following no attempt to correct SiPM arrays' gain  
for temperature excursions was done.  
Results with the calibration sources in situ 
are reported in table \ref{tab:calib}.
They are compatible, within errors, with the expected $\frac{1}{\sqrt{E}}$
behaviour. 
        
\begin{table}
\centering
\caption{Obtained FWHM resolution of Ce:LaBr$_3$ crystals with SiPM arrays as 
measured from calibration runs with \textsuperscript{57}Co and \textsuperscript{137}Cs sources obtained in situ at RAL.}
\smallskip
\begin{tabular}{ | l | c | c |c|}
\hline
used array & 59 keV   &  122 keV  &  662 keV \\
           & FWHM(\%) &  FWHM(\%) &  FWHM(\%) \\    
\hline
Hamamatsu  13361-AS (6 pcs)& $14.5 \pm 1.6$ & $9.6 \pm 1.4 $ & $5.0 \pm 0.9 $ \\
Hamamatsu 13361-AE (2 pcs) & $20.1 \pm 1.1$ & $13.4 \pm 3.1$ & $5.1 \pm 0.4 $ \\
Advansid ASD-P-4-TD  (1 pcs)     & 20 & 11.8 & 5.5 \\ \hline
\end{tabular}
\label{tab:calib}
\end{table}
Results from calibration in situ 
are a little worse than the ones obtained in laboratory.
This is due  to the presence of  environmental electronic noise, 
effects due to temperature  excursions, etc. 
 
Different readout arrays were used, including Hamamatsu S13361 SiPM arrays, 
made with a TSV  technology and Silicone (-AS) or epoxy (-AE) 
windows and Advansid  SiPM arrays. 

\begin{figure}
\centering
\includegraphics[width=8cm]{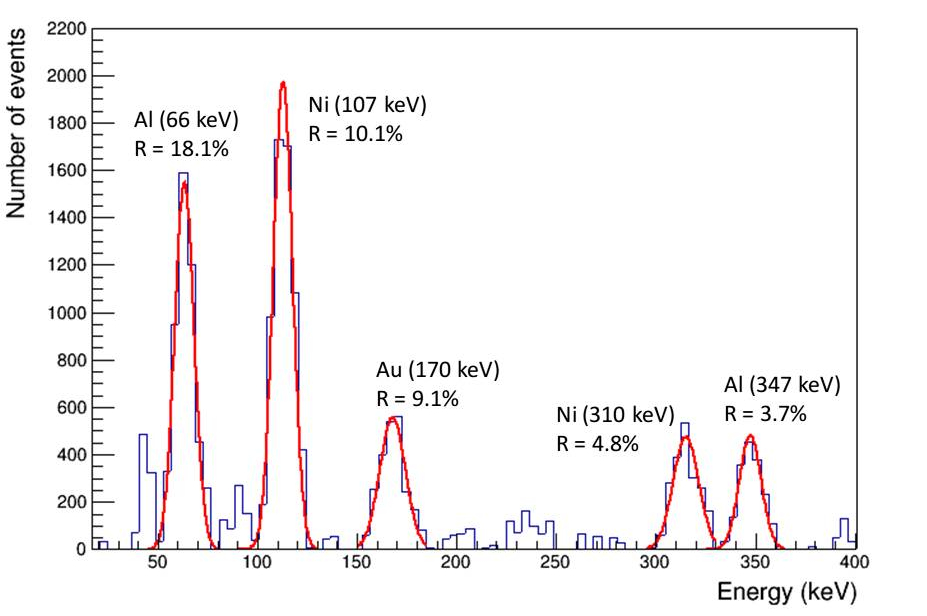}
\caption{Spectrum with a pure Hydrogen filling of the target, as seen from one
Ce:LaBr$_3$ detector, equipped with a Hamamatsu S13361-AS SiPM array. Background 
has been subtracted.}
\label{fig:back}
\end{figure}

Using a target filled with pure Hydrogen (background) the results of figure 
\ref{fig:back} were obtained. X-rays lines from  materials present in the 
target, such as Nickel, Gold, Copper and Aluminium are easily recognized. 

The detectors performance during data taking were then 
studied with delayed events (time larger than 2600 ns)
for a $H_2+O_2(0.3\%)$ mixture in the 2017 run. Unfortunately, the used gas
has been delivered by the supplier with an heavy contamination of $N_2$ and
thus X-rays lines from both Nitrogen and Oxygen
were present. Nevertheless, a spectral analysis was possible. For example a peak consistent with the $K_{\alpha}$ line from Nitrogen at 104.2 keV is reconstructed with a FWHM=11.3 keV, i.e. a resolution of 10.8 \%.


These preliminary results in a situation of a heavily polluted gas mixture, 
make us confident that compact 
1/2" Ce:LaBr$_3$ crystals with
SiPM array readout may be useful to instrument regions of 
the detector of difficult
access. As data were taken along periods with relevant temperature excursions,
the online correction of SiPM arrays' drift of gain with temperature will 
help to improve energy resolution up to the 4\% level (FWHM) obtained in 
laboratory  tests  in controlled environment \cite{bonesini17a}. 

\subsection{Detection of characteristic X-rays with HPGe detectors }
\begin{figure}[htbp]
\centering 
\includegraphics[width=1.2\textwidth]{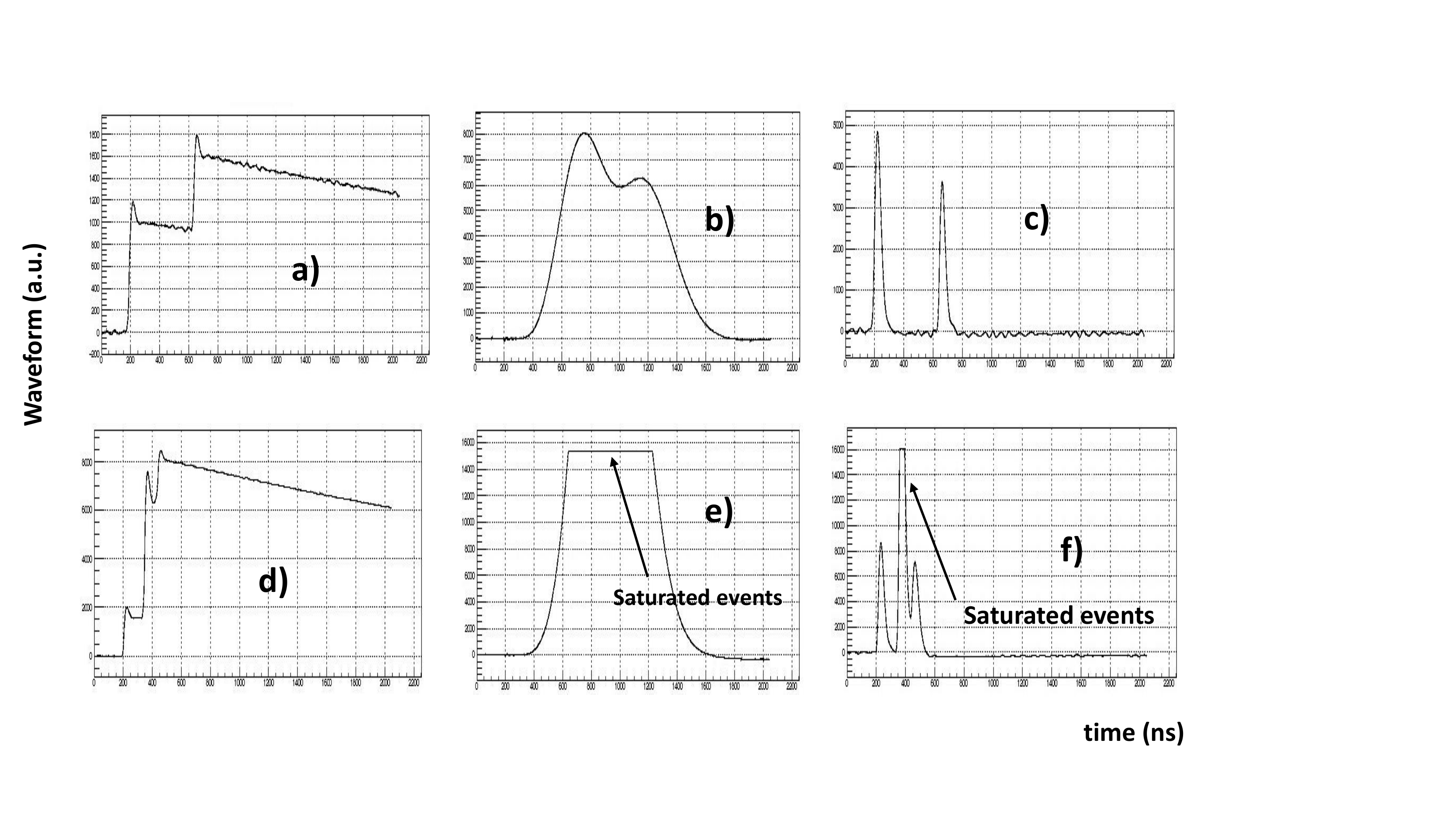}
\caption{(a),(d) pre-amplified signals for a HPGe GEM-S detector. 
(b),(c) amplified signals with
an Ortec 672 spectroscopy amplifier. (c),(f) amplified signals with a fast Ortec 579 shaping module.}
\label{fig:hpge1}
\end{figure}

The signal of HPGe detectors, after amplification and shaping, is characterized 
by a sharp risetime $\sim$100--300~ns
and a long tail $\sim$100--150~$\mu$s. 
A common limitation for the shaped signal in a high multiplicity environment
is the pile-up effect, where saturation may show up. 
Figure \ref{fig:hpge1} shows for a run taken with the 
$H_2+O_2(0.3 \%)$ target
two examples of signals from the GEM-S HPGe detector after 
the pre-amplification stage only, the amplification stage with a standard 
Ortec~672 spectroscopy amplifier (slow amplification, shaping time $2 \mu$s) 
and the amplification stage 
with a fast Ortec~579 amplifier (shaping time $\sim$200~ns). Two different situations are illustrated:
in the top panels a pile-up event is shown, while in the bottom panels a 
saturated event is shown. It can be seen that the use of a fast amplifier 
mitigates the presence of pile-up events.

\begin{figure}[htbp]
\centering 
\includegraphics[width=.69\textwidth]{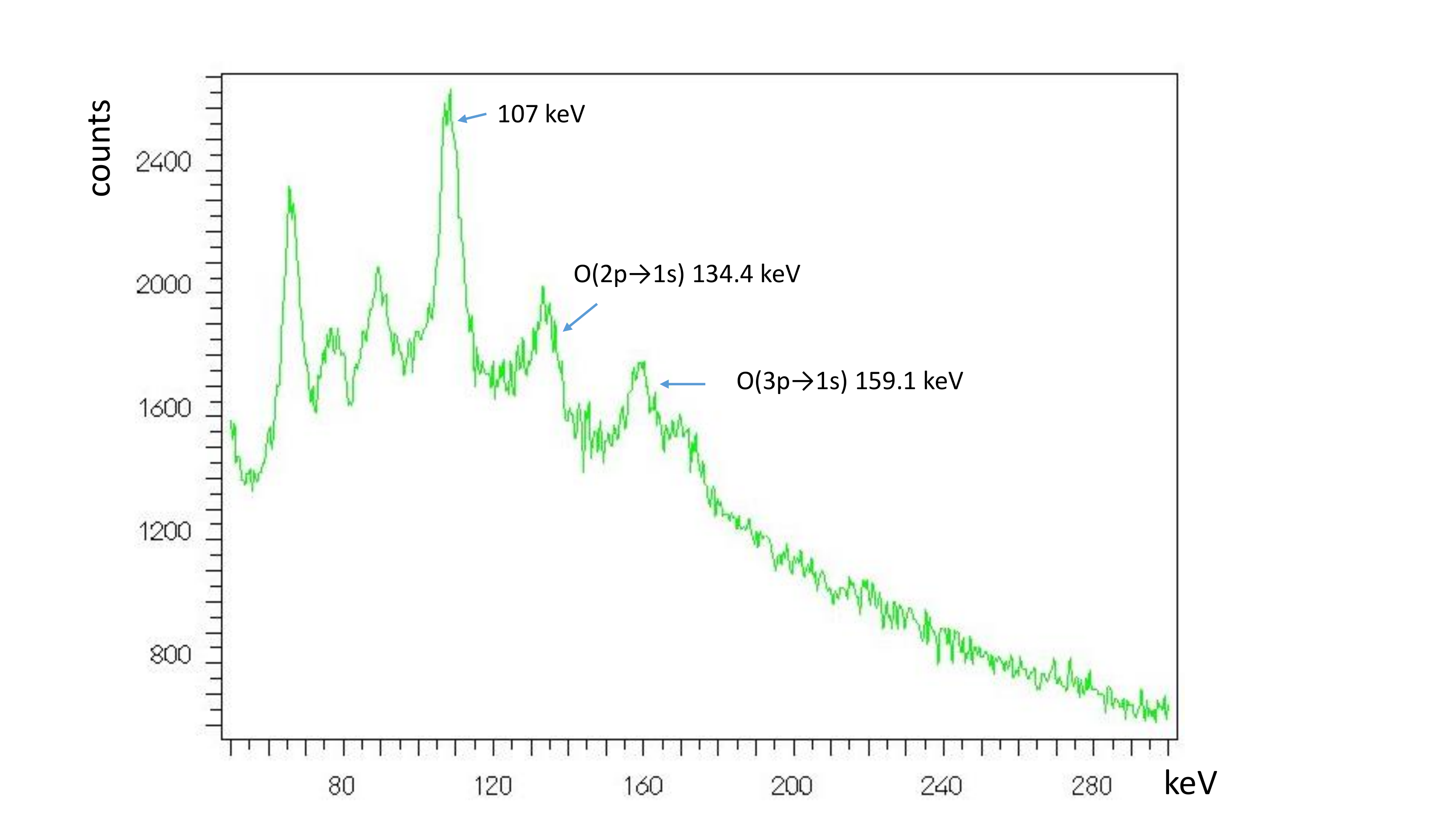}
\caption{50-300 keV energy spectrum as seen from the GEM-S HPGe detector
with a fast shaping of signals from an Ortec 579 amplifier.}
\label{fig:hpge2}
\end{figure}

The capability of HPGe detectors for intercalibration of the full apparatus is
demonstrated in figure \ref{fig:hpge2} for the HPGe GEM-S detector, 
with the  $H_2+O_2(0.3 \%)$ target, where the energy spectrum from
fast amplified signals with an Ortec~579 module are shown. 
Characteristic X-rays lines in the region
50--300~keV are clearly evident, while reconstruction results 
for the found X-ray lines are resumed in table 
\ref{tab2}.  
\begin{table}[!hbt]
    \caption{Characteristic X-lines as reconstructed by the GEM-S HPGe detector.}
  \begin{center}
\scalebox{0.8}{%
    \begin{tabular}{| l | l | c | c | c | c | c |c|}
      \hline
      & &  \multicolumn{2}{|c|}{pre amplifier} & \multicolumn{2}{|c|}{Ortec 672} & \multicolumn{2}{|c|}{Ortec 579}\\
      \hline
      transition    & Energy(keV) & E(keV) & FWHM(keV) & E(keV) & FWHM(keV) & E(keV) & FWHM(keV)  \\
      & (nominal)   & (rec.) &   & (rec.)  &  & (rec.) & \\
\hline
      Al $3d\rightarrow 2p$ & 66.1 & 65.9 & $4.5 \pm 0.1$ & 67.0 & $2.1 \pm 0.1$ & 66.3 & $5.5 \pm 0.1$ \\
      C $2p \rightarrow1s$ & 75.3 & 76.8 & $4.5 \pm 0.1$ & 77.2 & $2.1 \pm 0.1 $ & 77.2 & $5.5 \pm 0.1$ \\
      C $3p \rightarrow 1s$ & 89.2 & 88.9 & $4.5 \pm 0.1$ & 89.5 & $2.1 \pm 0.1 $& 89.2 & $5.5 \pm 0.1$ \\
Ni $4f \rightarrow 3d$ &107.4      & 107  & $5.1 \pm 0.4 $ &107 & $2.4 \pm 0.2$ & 107.3 &$5.5 \pm 0.1$ \\
      O $2p \rightarrow 1s$ & 133.5 & 133.9 & $5.4 \pm 0.5$ & 133.4 & $2.7 \pm 0.5$ &134.4 & $7.4 \pm 0.5$ \\
      O $3p \rightarrow 1s$ & 158.4 & 157.6 & $5.2 \pm 0.4$ & 157.5 & $3.5 \pm 0.5$ & 160 & $4.3 \pm 0.3$ \\
      O $4p \rightarrow 1s$ & 167.1 & - & - & - & - & 170.7 & $8  \pm 1$ \\
      Al $2p \rightarrow 1s$  & 346 & 347 & $6.7 \pm 0.2$ & 345 & $3.1 \pm 0.5$ & 348 &$12 \pm 1$\\                        
      \hline
    \end{tabular}}
  \label{tab2}
\end{center}
\end{table}

\begin{figure}[htbp]
\centering 
\vskip -0.3cm
\includegraphics[width=\textwidth]{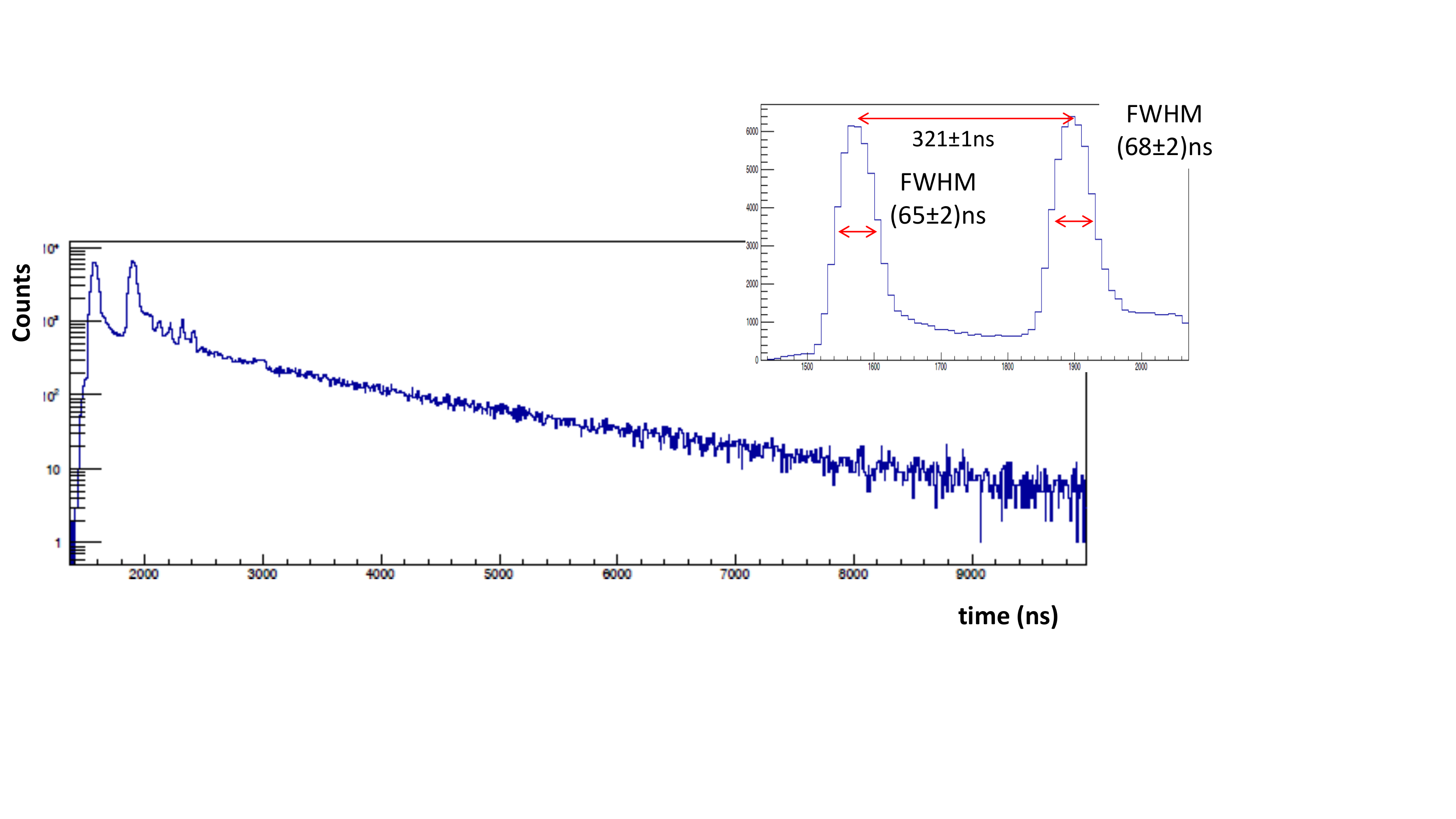}
\vskip -1.5cm
\caption{Reconstructed time structure of the beam spills from the GLP HPGe
detectors.}
\label{fig:hpge3}
\end{figure}

The timing properties of the HPGe detectors are instead shown 
in figure \ref{fig:hpge3}, for the GLP one. 
 The two peaks structure of the beam 
is shown in the enlarged inset, where the  time parameters for the two
beam spills in the beam: FWHM and distance between peaks are well reconstructed.  

\begin{figure}[htbp]
\centering 
\includegraphics[width=.89\textwidth]{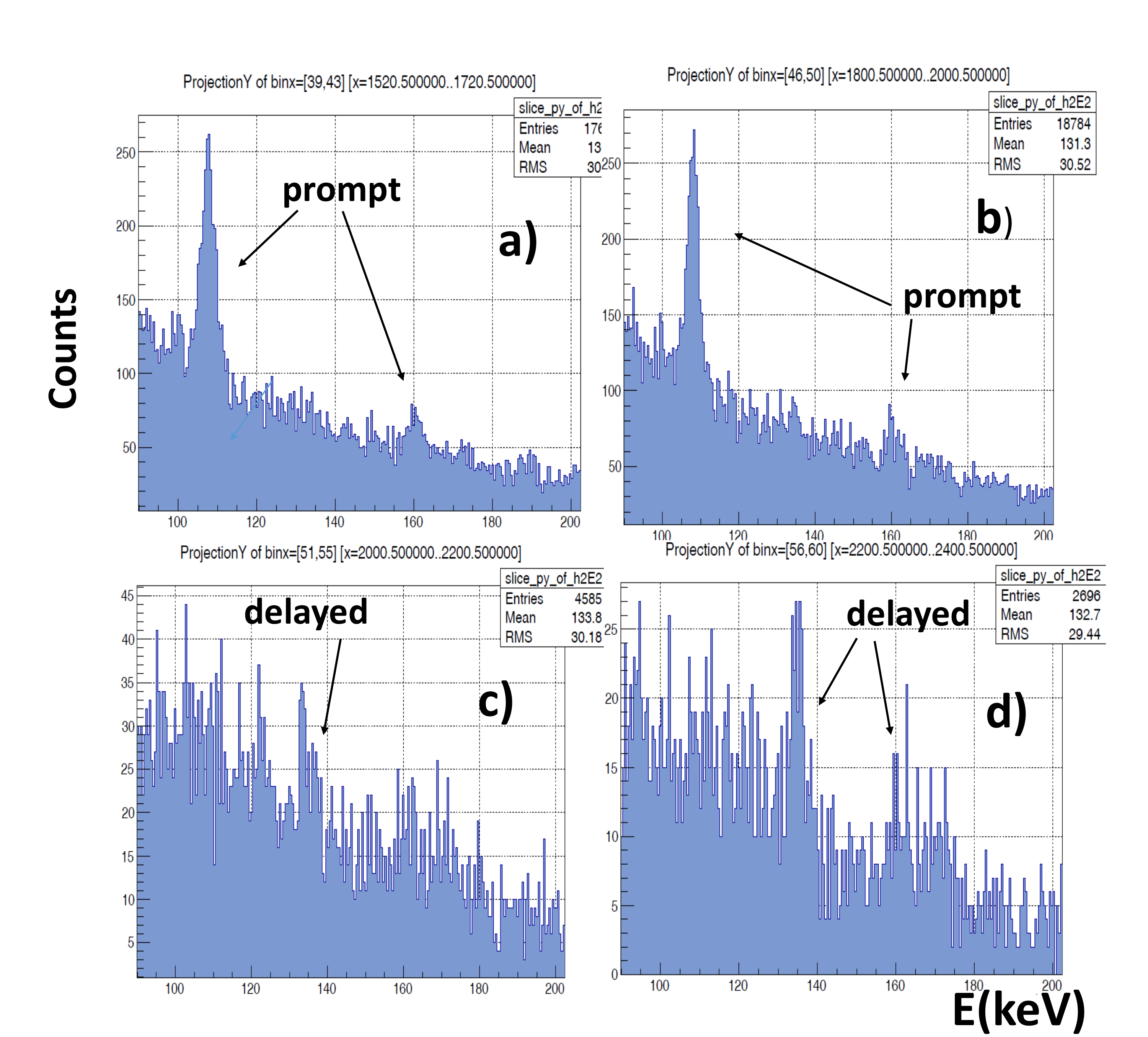}
\caption{Time evolution of the $K_{\alpha} (2p \rightarrow 1s) $ 
line energy distribution, as reconstructed from the GEM-S HPGe detector, in four
time slices. (a) 1520--1720~ns, (b) 1800--2000~ns, (c) 2000--2200~ns and 
(d) 2200--2400~ns.}
\label{fig:hpge5}
\end{figure}

The time evolution of the $K_{\alpha} (2p \rightarrow 1s)$ line  from
muonic Oxygen in events
recorded by the GEM-S HPGe detector for a $H_2+O_2 (0.3 \%)$ target, is shown 
in figure \ref{fig:hpge5}. Four successive time slices are shown: 1520--1720~ns
in panel (a), 1800--2000~ns in panel (b), 2000--2200~ns in panel (c) and last 
2200--2400~ns in panel (d).  

In conclusion, both types of HPGe detectors have shown in real high-rate and 
noisy running conditions, 
a good energy resolution
($\sim 2 \%$) in the 100~keV region of interest and a good time resolution and
are thus able to mitigate problems connected with pile-up events.

\section{Conclusions }
The FAMU experimental apparatus  has been used to study the muon transfer from 
muonic Hydrogen to the admixed high Z gases, as Oxygen or Argon.
An upgrade of this apparatus will
be used in the near future to measure the proton Zemach radius of the proton
with high precision by measuring the 1S muonic hydrogen hyperfine splitting. 
The setup used in the FAMU experiment at RAL has been described. 
It includes an high-pressure cryogenic target, a 1 mm pitch hodoscope for beam
characterization and a system of fast on purpose developed X-rays detectors,
based  on Ce:LaBr$_3$  crystals and HPGe detectors. 
Data have been recorded with a system based on CAEN FADC's in VME standard. 
Notwithstanding the high counting rate and background conditions, 
a clean detection of 
the muonic X-ray lines and a study  of their time evolution was possible.

\section*{Acknowledgements}
The research activity presented in this paper has been carried out in the 
framework of the FAMU experiment funded by Istituto Nazionale di 
Fisica Nucleare (INFN). The use of the low energy muons beam has been 
allowed by the RIKEN-RAL Muon Facility. We thank the staff of the mechanical
workshops of INFN Bologna, Milano Bicocca and Trieste for the continuos and
reliable support.  
We thank the RAL staff (cooling, gas, and 
radioactive sources sections) and especially Mr. Chris Goodway, Pressure 
and Furnace Section Leader, for their help, suggestions, professionalism 
and precious collaboration in the set up of the experiment at
RIKEN-RAL.

We thank Criotec srl and especially Ing. Adriano
  Mussinatto for the technical help and support in the costruction of
  the FAMU target.

A.~Adamczak and D. Bakalov acknowledge the support within the bilateral 
agreement between the Bulgarian Academy of Sciences and the Polish Academy 
of Sciences. D. Bakalov, P. Danev and M. Stoilov acknowledge the support 
of Grant 08-17 of the Bulgarian Science Fund.

We gratefully recognize the help of T. Schneider, CERN EP division,
 for his support in the optical cutting of the scintillating fibers of the
 hodoscope detector and the linked problematics,  N. Serra from
 Advansid srl  and M. Bombonati from Hamamatsu Photonics Italia srl for useful 
discussions on SiPM  and A. Abba from Nuclear Instruments srl for aid
 in the problematics of temperature control of SiPM.

\end{document}